\documentclass[preprint,amsmath]{aastex}

\newcommand{\etal}{{\it et al. }}
\newcommand{\mas}{\mu\mathrm{as}}
\newcommand{\be}{\begin{equation}}
\newcommand{\ee}{\end{equation}}

\slugcomment{to appear in PASP, March 2006}
\shorttitle{Multiple Companions... SIM Planet Searches}
\shortauthors{Ford}

\begin{document}
\title{The Effects of Multiple Companions on the Efficiency of the SIM Planet Searches}
\author{Eric B.\ Ford\altaffilmark{1} }

\affil{Astronomy Department,
	601 Campbell Hall, 
	University of California at Berkeley, 
	Berkeley, CA 94720-3411, USA}
\affil{Department of Astrophysical Sciences, 
	Princeton University, 
	Peyton Hall, 
	Princeton, NJ 08544-1001, USA}
\email{eford@astron.berkeley.edu}

\altaffiltext{1}{Miller Research Fellow}

\begin{abstract}
The Space Interferometry Mission (SIM) is expected to make precise
astrometric measurements that can be used to detect low mass planets
around nearby stars.  Since most nearby stars are members of multiple
star systems, many stars will have a measurable acceleration due to
their companion, which must be included when solving for astrometric
parameters and searching for planetary perturbations.  Additionally,
many of the stars with one radial velocity planet show indications of
additional planets.  Therefore, astrometric surveys like SIM must be
capable of detecting planets and measuring orbital parameters in
systems with multiple stellar and/or planetary companions.  We have
conducted Monte Carlo simulations to investigate how the presence of
multiple companions affects the sensitivity of an astrometric survey
such as SIM.  We find that the detection efficiency for planets in
wide binary systems is relatively unaffected by the presence of a
binary companion, if the planetary orbital period is less than half
the duration of the astrometric survey.  For longer orbital periods,
there are significant reductions in the sensitivity of an astrometric
survey.  Additionally, we find that the signal required to detect a
planet can be increased significantly due to the presence of an
additional planet orbiting the same star.  Fortunately, adding a
modest number of precision radial velocity observations significantly
improves the sensitivity for many multiple planet systems.  Thus, the
combination of radial velocity observations and astrometric
observations by SIM will be a particularly valuable for studying
multiple planet systems.
\end{abstract}


\section{Introduction}

During the last decade, high precision radial velocity surveys have
discovered $\sim$170 planets orbiting nearby stars (Butler \etal 2002 and
references therein; \url{http://exoplanets.org};
\url{http://www.obspm.fr/planets}).  Many of the planets orbit stars
believed to be members of wide binary systems ($\sim1000$ AU;
Eggenberger, Udry, \& Mayor 2004), and nearly $\sim20$ of the stars
with one radial velocity planet have at least one additional
companion.  Additionally, there are indications that many more stars
with planets show residuals suggestive of additional planets (Fischer
\etal 2001).

Astrometric planet searches could provide valuable information about
nearby planetary systems and planets in multiple star systems.  While
radial velocity measurements are only capable of measuring $m \sin i$,
the mass of a planet times the inclination of its orbital plane to the
plane of the sky, astrometric observations measure both the mass and
inclination separately.  Accurate masses and inclinations are
important for dynamical analyses of multiple planet systems.  (e.g.,
Ford, Lystad, \& Rasio 2005; Lee \& Peale 2002).  For example,
classical planet formation theory predicts that planets form on nearly
coplanar orbits.  Astrometric measurements of multiple planet systems
could test this prediction.

Similarly, astrometric planet searches could yield insight into possible
differences in the planets around single and binary
stars.  For example, Zucker \& Mazeh (2002) claim to detect a
mass-period correlation and that it is different for planets around single
stars and planets around one star in a binary system.  There are also
theoretical reasons to believe that planetary systems may be affected
in binary systems.  For example, it is now believed that dynamical
interactions between a planet and a wide binary companion can result
in significant evolution of a planet's orbit.  In systems such as 16
Cygni, the distant stellar companion (16 Cyg A) may be responsible for
large oscillations in the orbital eccentricity of a planet around 16
Cyg B (Holman, Touma, \& Tremaine 1997).
More speculatively, some have proposed that distant binary companions
might even produce giant planets with orbital periods of $\sim1-4$ d,
that are difficult to produce in conventional theories of planet
formation.  If the orbital plane of the stellar binary and the orbital
plane of the planet are nearly perpendicular, then secular
interactions can result in very large eccentricity oscillations and a
planet reaching a periastron of only a few stellar radii.  In such a
case, tidal dissipation might result in a planet's orbit becoming
circularized with a semi-major axis of several stellar radii (Ford,
Rasio, \& Sills 1998; Faber, Rasio, \& Willems 2004; Takeda \& Rasio 2005).  

The Space Interferometry Mission (SIM) will make precision astrometric
measurements suitable for detecting planets around nearby stars (Marcy
\etal 2005; Sozzetti 2005).  Two of the key projects are surveys to
search for low-mass planets (Marcy \etal 2002; Shao \etal 2002), many
of which are likely to be in multiple planet systems or wide stellar
binary systems.  Therefore, it is important to understand how the
sensitivity of an astrometric planet search is impacted by the
presence of multiple planets and wide binary systems.  
Additionally, SIM is expected to conduct follow up observations of some 
planetary systems discovered by radial velocity surveys.  If giant
planets serve as tracers of planetary systems containing low-mass
planets, then these follow up observations could also yield new
detections of low-mass planets in planetary systems also containing
one or more giant planets.  Given that such solar-system analogues
would be very exciting, it is important to understand the sensitivity
of SIM to multiple planet systems.  

Similarly, since we know that most nearby stars are members of binary
star systems, it is important to understand how a wide binary
companion would affect the sensitivity of SIM to planetary companions.
If the acceleration due to a distant companion were to interfere with
SIM's ability to detect low mass planets, then it might be prudent to
remove stars with binary companions from the SIM target list in favor
of more distant single stars.  Alternatively, if SIM is able to detect
low mass planets equally well around single stars and stars in wide
binary systems, then it would be desirable to target the closest stars
whether or not they are in binaries.  Targeting both single and binary
stars could provide information about how the formation of planetary
systems differs between these two types of stars.
Previously, Sozzetti \etal (2002) and Ford \& Tremaine (2003) have
investigated the sensitivity of SIM for detecting a single planet
around a star and measuring its mass and orbital parameters.
Additionally, Sozetti \etal (2003) have begun to investigate the
potential of SIM for detecting and characterizing multiple planet
systems.  They concentrate on giant planets and use the known radial
velocity systems as a guide.  This study considers SIM planet searches
for two-planet systems more generally.  In particular, we focus on how
the sensitivity of SIM to lower mass planets is affected by the
presence of an addition planet.
%
%

In this paper, we explore the sensitivity of astrometric surveys for
detecting and characterizing the orbits of planets in multiple planet
systems or in wide binaries.  While we have chosen simulation
parameters to approximate the planned SIM narrow-angle planet
searches, many of our simulation results can be rescaled for various
measurement precisions, planet masses, and stellar distances which
can be of use for other astrometric planet searches, such as those
being planned for PRIMA and GAIA.  Further, our conclusions should
be qualitatively applicable to most astrometric planet searches.  We
summarize our assumptions and methods in \S 2.  In \S 3 we present the
results of our simulations for planets in binaries systems.  In \S 4
we describe the particular model for planets in multiple planet
systems and address several specific questions regarding the discovery
potential of SIM for two-planet systems.  In \S 5, we show how the
addition of radial velocity measurements can increase the sensitivity
of SIM to certain types of multiple planet systems.  Finally, we
summarize our main findings and conclusions in \S 6.

\section{Methods}

We simulate astrometric observations following the method of Ford
(2004).  For each star we simulate $2 N_{obs} = 48$ (for $P_{SD} = 5$
years) or $2 N_{obs} = 96$ (for $P_{SD} = 10$ years) one dimensional
astrometric observations grouped in pairs.  The $N_{obs}$ observation
times ($t_i$) are distributed over the survey duration $P_{SD}$.  Each
pair of observations occurs nearly simultaneously and is made with
perpendicular baselines.  A Gaussian noise source with standard
deviation $\sigma_{SIM}$ is added to each one dimensional observation.
We used the ``periodic with perturbations'' observing schedule (with
$\epsilon=0.2$) discussed in Ford (2004).  In this schedule the
observation times are periodic, except that Gaussian perturbations
with standard deviation $\epsilon P_{SD} / N_{obs} \simeq 15$ d have
been applied to the time of each observation, to reduce the effects of
aliasing near the orbital period $P_{SD}/N_{obs} = 76$ d.  While the
simulations of Ford (2004) demonstrated the merits of this observing
schedule, that study was restricted to observing schedules that were
not able to use information obtained from early observations for the
scheduling of future observations.  

One way in which SIM's eventual sensitivity might be improved is by
the application of adaptive scheduling algorithms that use of early
observations to improve the scheduling of later observations, such as
proposed by Loredo (2004).  Ford (2005b) developed computationally
feasible algorithms for radial velocity surveys and demonstrated that
adaptive scheduling algorithms could significantly improve the
efficiency of radial velocity planet searches.  It would be
straightforward (but computationally demanding) to implement similar
adaptive scheduling algorithms for astrometric surveys, potentially
increasing the efficiency of SIM planet searches.  Due to the larger
number of parameters required to describe the astrometric motion of an
isolated star (position, parallax, and proper motion as compared to
only the line-of-sight radial velocity), we expect that a somewhat
larger number of early observations will be necessary before the
benefits of adaptive scheduling become significant.  We expect that
future studies will implement adaptive scheduling algorithms for
astrometric searches and evaluate their performance characteristics.
In this study, we restrict our attention to fixed observing schedules,
primarily due to the demanding computational requirements of such
adaptive scheduling algorithms, especially given that we must simulate
millions of planetary systems to explore the very large parameter
space necessary to describe multiple planet systems.

For SIM's narrow angle planet surveys, each ``single measurement'' of
a target star will consist of multiple measurements of the optical
path length delay for the target star and multiple nearby reference
stars while maintaining the spacecraft orientation.  Each pair of
optical path length delay measurements for two different stars results
in a relative delay measurements between the target star and a
reference stars.  This one dimensional relative delay measurement is
proportional to the the angular distance between the two stars
projected onto the interferometer baseline (Marcy \etal 2002; Shao
\etal 2002).  The single measurement precision that we refer to is the
precision obtained by combining multiple one dimensional relative
delay measurements between the target star and multiple reference
stars in a single visit (~1 hour).  In our simulations, we set
$\sigma_{SIM} = 1 \mas$, but most of our results can be scaled to
other values of $\sigma_{SIM}$ using the ``scaled signal'',
\be
S \equiv \frac{\alpha}{\sigma_{SIM}},
\ee
where 
\be
\frac{\alpha}{''} \equiv \frac{m}{M} \frac{a}{\mathrm{AU}} \frac{\mathrm{pc}}{D},
\ee
$\alpha$ is the semi-amplitude of the stars motion on the plane of the sky,
$m$ is the mass of the planet, $M$ is the mass of the star, $a$ is the semi-major
axis of the planet, and $D$ is the distance to the star.  
For example, a Jupiter mass planet ($M_J$) with a 3d, 1yr, or 5yr
orbital period would result in $S=3.9$, $96$, or $280$, assuming a 1
$M_\odot$ star at a distance of $10$ pc.  Similarly, a Neptune (Earth)
mass planet orbiting a 1 $M_\odot$ star $10$ pc away would induce a
perturbation of magnitude $S=0.21$, $5.1$, or $15$ (0.012, 0.30, or
0.88) for a 3d, 1yr, or 5yr orbital period.

Stars in wide binary systems will have an unknown acceleration in
addition to the usual astrometric parameters (position, distance, and
proper motion).  The angular acceleration ($\vec{a}$) induced by a
wide binary companion is 
$$\frac{\vec{a}}{\mu\mathrm{as}\, \mathrm{yr}^{-2}} = 4 \left(\frac{M_c}{M_\oplus}\right) \left(\frac{r}{1000 \mathrm{AU}}\right)^{-2} \left(\frac{D}{10 \mathrm{pc}}\right)^{-1} \cos i,$$
where $M_c$ is the mass of the binary companion, $r$ is the three
dimensional separation between the primary and its companion, $D$ is
the distance between the target star and the Earth, and $\cos i$ is
the cosine of the angle between the plane of the sky and the vector
between the two stars.  This can be easily obtained from Newton's laws
by solving for the acceleration of a body due to due to a point mass
($M_c$) at distance ($r$), projecting onto the plane of the sky, and
dividing by the distance to the star ($D$).  We restrict our attention
to wide binary companions for which the orbital period is much longer
than the duration of the astrometric survey (the expected SIM mission
lifetime is 5-10 years).  Thus, the perturbations due to the binary
companion can be accurately modeled as a constant acceleration of the
target star.

Stars with multiple planetary mass companions will have an astrometric
signal that is the superposition of the perturbations from each
planet, implicitly depending on the gravitational perturbations
between the planets.  In this paper, we model the motion of each
planet as a Keplerian orbit, even for systems with multiple planets.
While gravitational interactions among the planets can result in
significant deviations from purely Keplerian orbits (e.g., GJ 876),
most of the planetary systems discovered by the radial velocity method
can be well modeled by planets on non-interacting Keplerian orbits, at
least for time scales comparable to the expected lifetime of SIM.

\subsection{Model Planets and Observations}

We simulate astrometric observations of many hypothetical stars.  Each
star is assigned a position ($RA$, $Dec$), distance ($D$), proper
motion ($\vec{v_\perp}/D$), mass ($M$).  The stellar positions are
distributed uniformly in a shell of radius 10 parsecs centered on the
Sun.  The stellar velocities are drawn from a three-dimensional
Gaussian distribution with mean $0$ km s$^{-1}$ and standard deviation
$40$ km s$^{-1}$ in each direction.  The stellar masses are set to $1
M_\odot$ and the stellar velocities are randomly directed in space.
When considering planets in wide binary systems in \S 3, we augment
the model of Ford (2004) by assigning each star an angular
acceleration ($\vec{a}$).  Each star is also assigned one or two planets,
depending on whether considering the effects of wide stellar binaries
in \S 3 or multiple planet systems in \S 4.

As in Ford (2004), each planet is assigned a mass ($m$), orbital
period ($P$), orbital eccentricity ($e$), inclination of the orbital
plane to the plane of the sky ($i$), argument of pericenter
($\omega$), longitude of ascending node ($\Omega$), and mean anomaly
at a specified time ($M_o$).  The masses and orbital periods of the
simulated planets are drawn independently from the planetary mass and
period function determined by Tabachnik \& Tremaine (2002) and are
based on radial velocity observations of massive planets.  The planet
masses range from one Earth mass ($1 M_\oplus$) to ten Jupiter masses
($10 M_{Jup}$) and the orbital periods range from $2$ days to a
maximum orbital period, $P_{\max} = 2 P_{SD}$, twice the duration of the
astrometric survey ($P_{SD}$).  The eccentricities are drawn from a
uniform distribution between $0$ and $1$.  The orbits are drawn
independently and randomly oriented in space.
%
%
We consider both $P_{SD} = 5$
and $10$ years, since there can be significant differences in the
efficiency of detection for planets in the habitable zone depending on
the duration of the survey (as we will show in \S 4).  For two-planet
systems, we have conducted additional simulations fixing the mass
and/or orbital period of one or both planets to help visualize how
the sensitivity depends on the other parameters.

\subsection{Data Analysis}

After simulating the observations, we attempt to fit a no-planet model
that includes only the star's five astrometric parameters ($D$, $RA$,
$Dec$, and the two components of $\vec{v}$) as in Ford (2004).  In all
our fitting we use use the Levenberg-Marquardt algorithm (Press \etal
1992) combined with good initial guesses of the parameters as
described in Ford (2004).  When considering the effects of planets in
binary star systems in \S 3, we also allow the no-planet model to
include an unknown acceleration.  After identifying the best-fit
astrometric parameters, we evaluate the appropriateness of the
no-planet model by a $\chi^2$ test using the usual sum of squares of
the residuals for the no-planet model $\chi_0^2$.

When considering the effects of planets in binary star systems, if the
no-planet model (allowing for an unknown constant acceleration) can be
rejected with 99.9\% confidence, then we apply the Levenberg-Marquart
algorithm to obtain the best-fit model with both a constant
acceleration and one planet.  When fitting for orbital parameters, we
use use Thiele-Innes coordinates as described in Ford (2004).  We
calculate $\chi_1^2$ for the new model (including both a constant
acceleration and one planet) and detect the planet if $\chi_{1}^2$ is
significantly less than $\chi_0^2$, according to an $F$-test.

When considering two-planet systems in \S 4 \& 5, if a $\chi^2$ test
can reject the no-planet model with 99.9\% confidence, then we proceed
to fit two new models, each including only one of the two planets
(labeled a \& b).  We calculate $\chi_{1a}^2$ and $\chi_{1b}^2$ for
these two models and detect planet a (b) if $\chi_{1a}^2$
($\chi_{1b}^2$) is significantly less than $\chi_0^2$, according to an
$F$-test.  If both one-planet models can be rejected with 99.9\%
confidence by a $\chi^2$ test, then we proceed to fit a two-planet
model (including both planets a \& b) and calculate $\chi_2^2$.  If
$\chi_2^2$ represents a significant improvement from both
$\chi_{1a}^2$ and $\chi_{1b}^2$ according to an $F$-test, then we
detect both both planets (a \& b) simultaneously.  Note that it is
technically possible to detect both planets simultaneously but not
individually (if $\chi_{2}^2$ is significantly better than
$\chi_{0}^2$, even when neither $\chi_{1a}^2$ nor $\chi_{1b}^2$ are).
It is also technically possible to detect both planets individually
but not simultaneously (if both $\chi_{1a}^2$ and $\chi_{1b}^2$ are
significantly better than $\chi_{0}^2$, but $\chi_{2}^2$ is not).
This can occur when the data provide evidence for at least one planet,
the data still permit multiple very different orbital solutions, and
the data do not yet provide sufficient evidence for two planets.  In
our subsequent discussion of two-planet systems, we consider planet a
(b) to be detected if either $\chi_{1a}^2$ ($\chi_{1a}^2$) or
$\chi_{2}^2$ is significantly less than $\chi_0^2$ according to an
$F$-test.  This criteria is slightly optimistic since it occasionally
counts a planet as detected even when multiple qualitatively different
orbital solutions are still viable models.

While Bayesian techniques have already been applied to orbit fitting
(Ford 2005a) and model selection (Ford 2005b) for radial velocity
planet surveys, all previous studies of SIM planet searches have
relied on frequentist statistical methods.  We recognize the
superiority of a Bayesian approach and expect that future work will
develop the algorithms necessary for Bayesian parameter estimation and
model selection to be applied to the actual SIM planet searches.  Here
we employ frequentist methods due to computational limitations, since
we must simulate orders of magnitude more systems than will actually
be observed by SIM.

We repeat these calculations for millions of stars to determine the
efficiency for detecting planets with in various binary star and
multiple planet systems.  It is important to note that, when finding
best-fit models, we use only a local search algorithm, primarily due
to the computational limitations when performing millions of orbital
fits.  We use parameters for both the star and planet typically
accurate to 1\%.  Our results should be considered optimistic.  In
practice, it will be necessary to conduct a global search to provide
initial guesses and it may be more difficult for a global search to
find the best-fit model.  This is a subject worthy of separate
investigation.  While Konacki, Maciejewski \& Wolszczan (2002) provide
one method of obtaining initial guesses, it has not been demonstrated
to work for multiple planet systems or for systems with a modest
signal-to-noise ratio.

\section{Results for Planet in Binary Star Systems}

We have simulated observations of many stars with a planet to
determine our ability to detect planets and measure their masses and
orbital parameters.  To understand the effects of a wide binary
companion, we compare the results of simulations in which we assume an
unknown acceleration term with simulations in which we assume the
acceleration is known.  We also compare the results of simulations
with various values for the magnitude of the acceleration, the
planetary orbital period, the planetary eccentricity, and the
magnitude of the scaled signal.

\subsection{Rates versus Acceleration}

\begin{figure}[ht]
\plotone{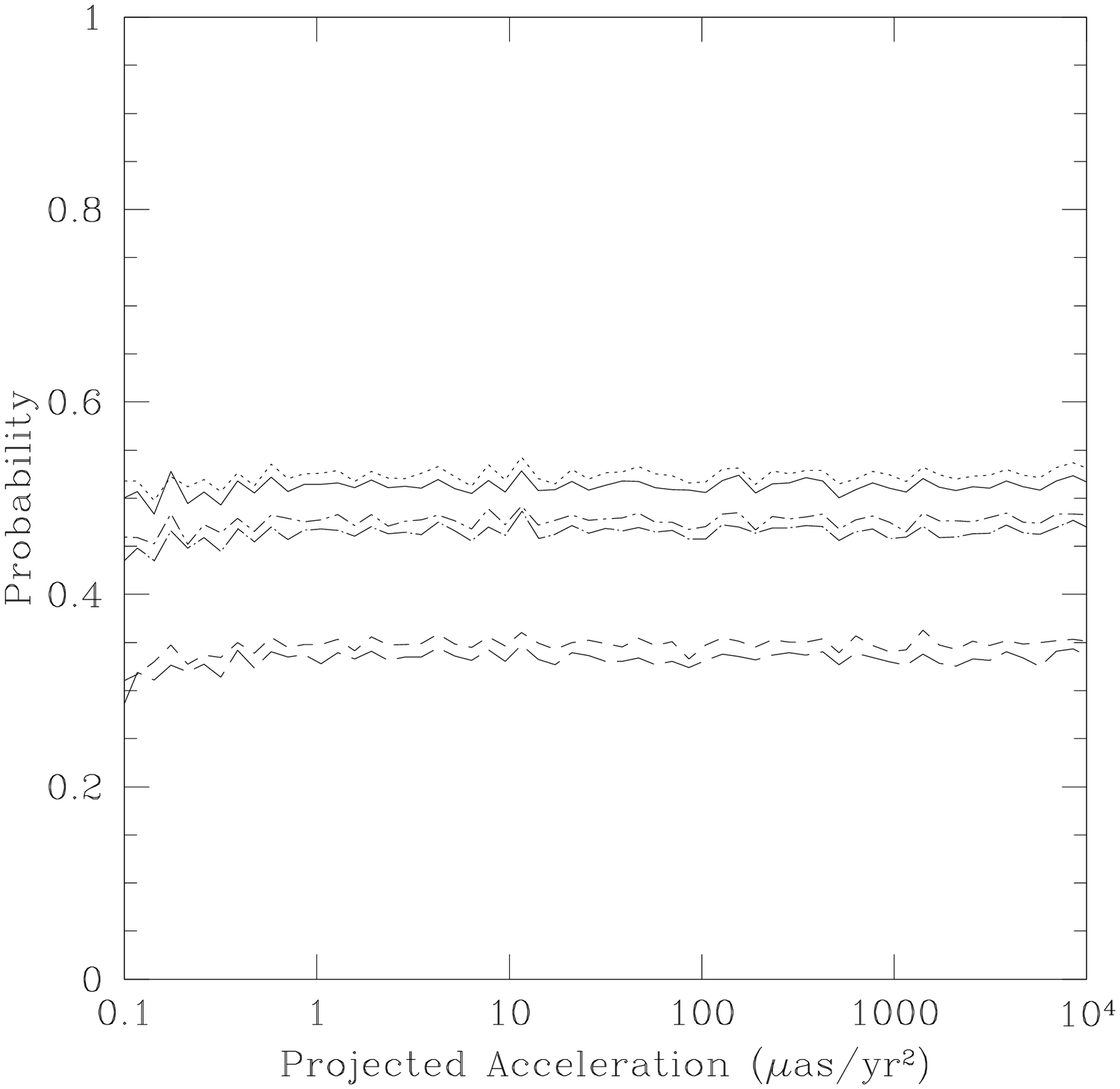} 
\caption[f1.eps]{
Here we compare the overall rates for detecting and characterizing
planets as a function of the projected angular acceleration, averaged
over the distance to the star, planetary mass, planetary orbital
period, and other parameters.  The top pair of lines is for
detections, the middle pair of lines is for measuring the planetary
mass with 30\% accuracy, and the bottom pair of lines is for
measuring the planetary orbital parameters with 10\% accuracy.  In
each pair the top line is for stars with a known acceleration and
the bottom line is for stars with an unknown acceleration.  The additional
acceleration has a negligible effect on the overall efficiency.
\label{FigRateVsAcc}}
\end{figure}

First, we compare the overall rates for detecting and characterizing
planets as a function of the projected angular acceleration, averaged
over the distance to the star, planetary mass, planetary orbital
period, and other parameters ($e$, $\omega$, $i$, $\Omega$, and $M_0$;
see Fig.\ \ref{FigRateVsAcc}).  The top pair of lines is for
detections, the middle pair of lines is for measuring the planetary
mass with 30\% accuracy, and the bottom pair of lines is for measuring
the planetary orbital parameters with 10\% accuracy.  In each pair the
top line is for stars with a known acceleration and the bottom line is
for stars with an unknown acceleration.  In each case, the unknown
acceleration slightly reduces the overall rates for detecting and
characterizing the planet, but the size of the effect is independent
of the magnitude of the acceleration.  Therefore, for the remainder of
the paper, we set the actual acceleration term to zero, and compare
the results obtained by assuming that the acceleration is known to be
zero to the results when we consider the acceleration to be unknown
and attempt to find the best-fit acceleration term.

\subsection{Rates versus Scaled Signal}

\begin{figure}[ht]
\plotone{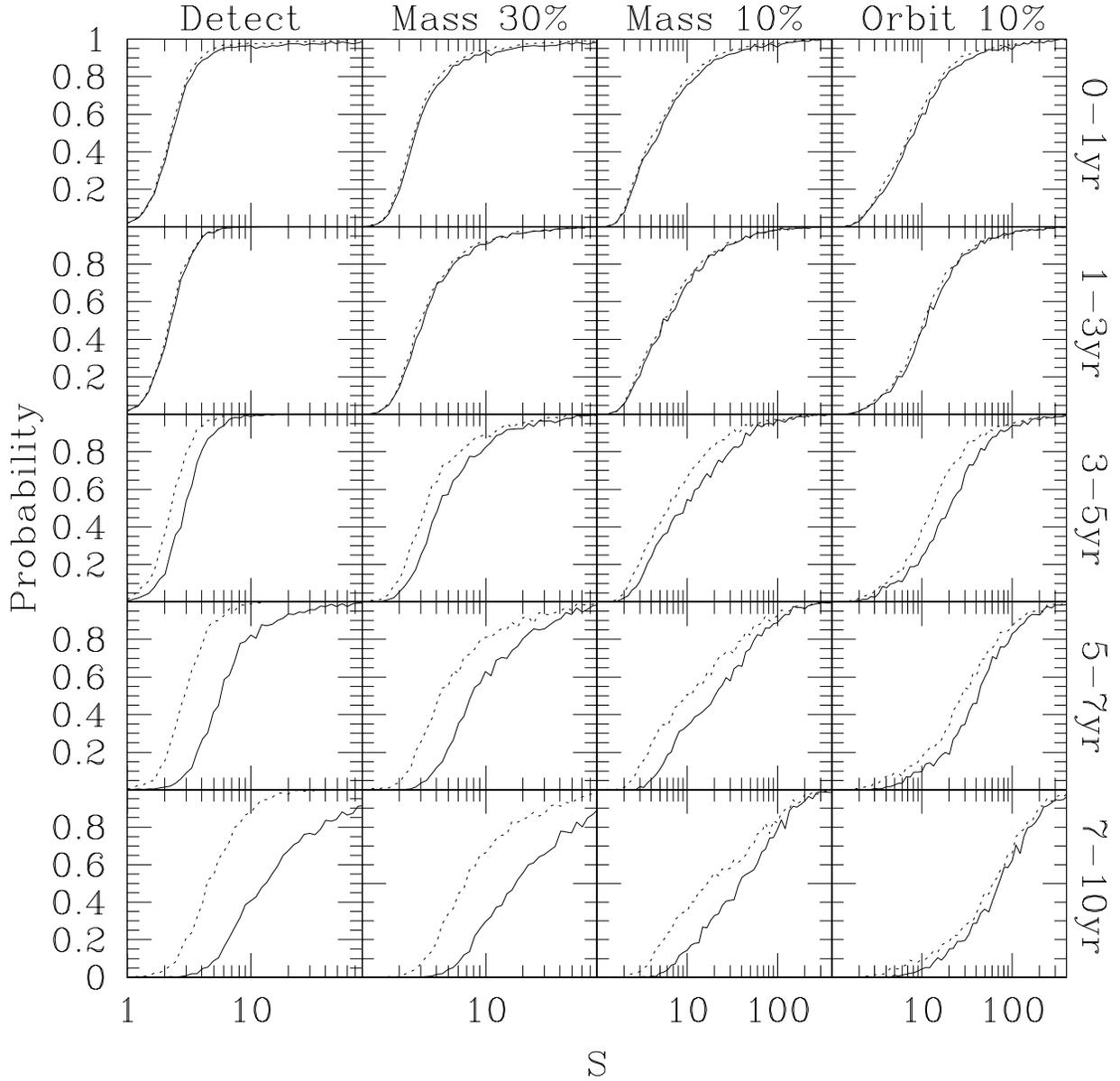} 
\caption[f2.eps]{
Here we plot the rates for detecting and characterizing planets as a
function of the scaled signal, both including (solid lines) and
omitting (dotted lines) a constant acceleration term.  Each row of
panels is for a different range of planetary orbital periods (top row:
0-1 yr, second row: 1-3 yr, middle row: 3-5 yr, fourth row: 5-7 yr,
bottom row: 7-10 yr).  The mission duration is assumed to be 5 yr.
The far-left column of panels is for detecting
the planet, the center-left column is for measuring the mass with 30\%
accuracy, the center-right column is for measuring the mass with 10\%
accuracy, and the far-right column is for measuring the orbital
parameters with 10\% accuracy.
\label{FigRateVsS}}
\end{figure}

\begin{figure}[ht]
\plotone{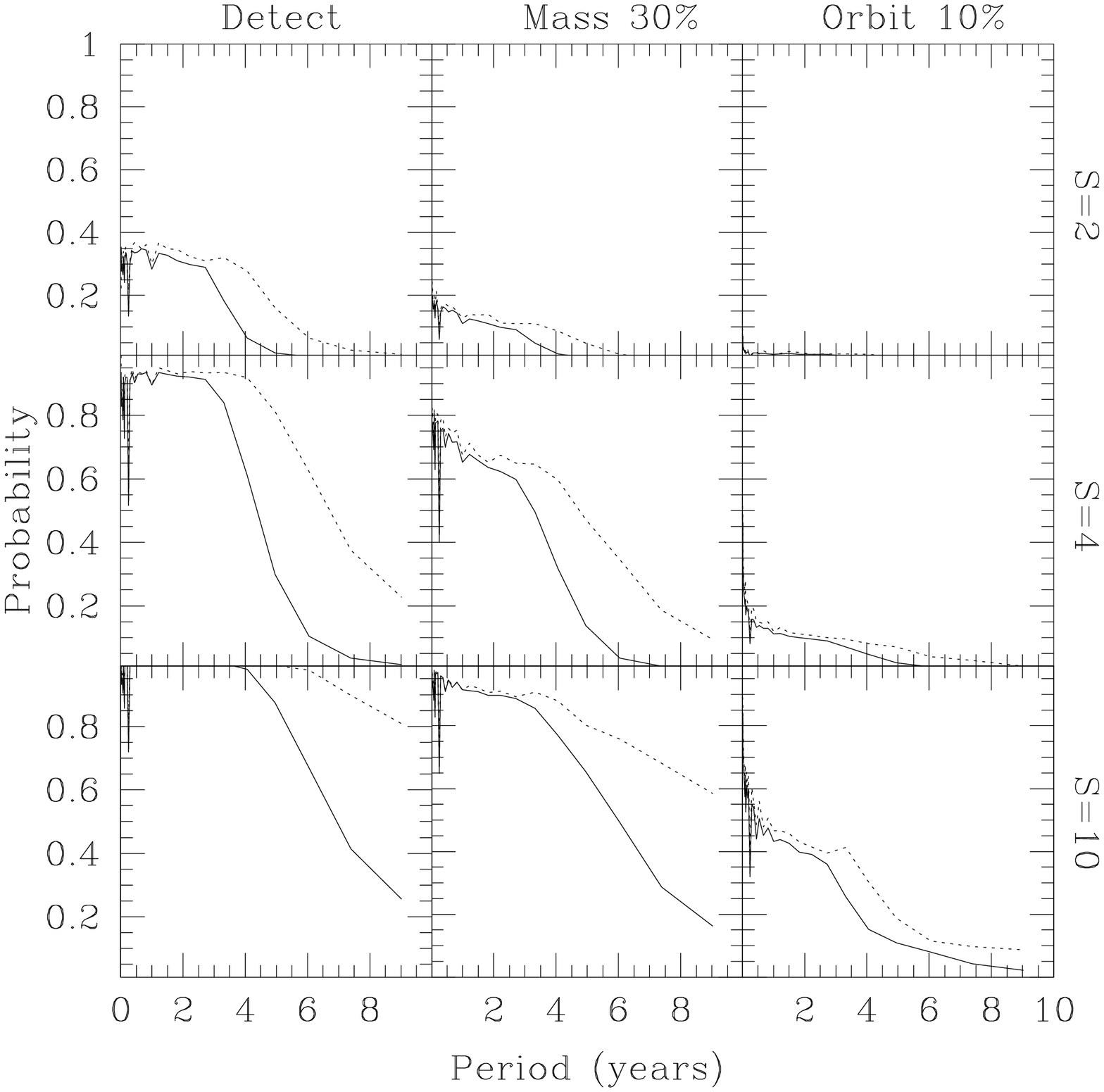} 
\caption[f3.eps]{
Here we show the rates for detecting and characterizing planets as a
function of the planetary orbital period.  In each panel the solid
line is for models including an unknown constant acceleration and the
dotted line assumes no acceleration.  Each row of panels is for a
different fixed value of the scaled signal (top: $S=2$, middle: $S=4$,
bottom: $S=10$).  The left column of panels is for detecting the
planet, the center column is for measuring the mass with 30\%
accuracy, and the right column is for measuring the orbital
parameters with 10\% accuracy.
\label{FigRateVsPConstS}}
\end{figure}

\begin{figure}[ht]
\plotone{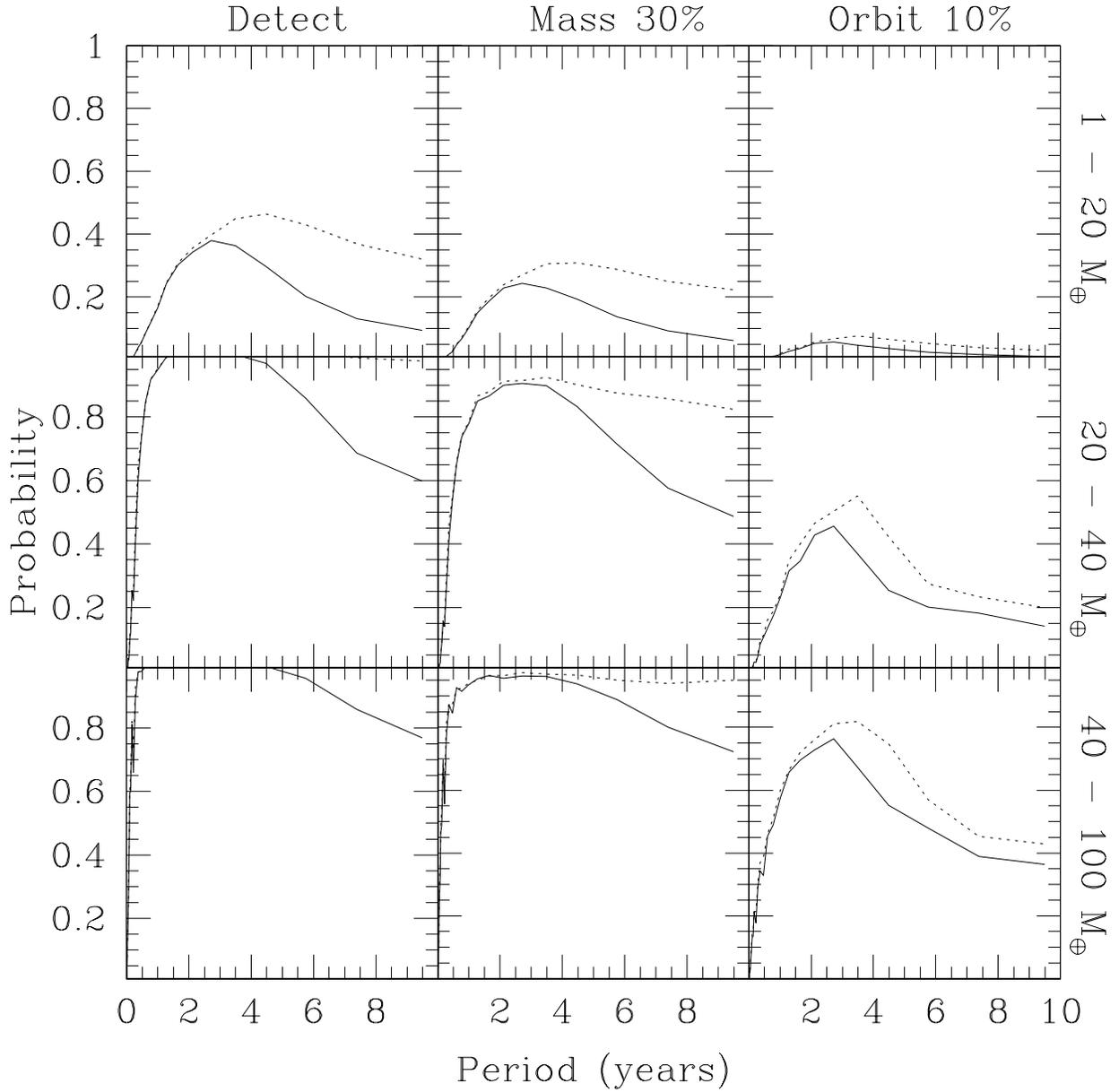} 
\caption[f4.eps]{
Here we show the rates for detecting and characterizing planets as a
function of the planetary orbital period.  In each panel the solid
line is for models including an unknown constant acceleration and the
dotted line assumes no acceleration.  Each row of panels is for a
different range of planetary masses (top: $1-20 M_\oplus$, middle:
$20-40 M_\oplus$, bottom: $40-100 M_\oplus$).  The left column of
panels is for detecting the planet, the center column is for measuring
the mass with 30\% accuracy, and the right column is for measuring
the orbital parameters with 10\% accuracy.
\label{FigRateVsPConstM}}
\end{figure}

While Fig.\ \ref{FigRateVsAcc} demonstrates that the overall rates for
detecting and characterizing planets is only slightly affected by
including an unknown constant acceleration, the effect can be much
more significant for certain values of the scaled signal and planetary
orbital period.
In Fig.\ \ref{FigRateVsS} we plot the rates for detecting and
characterizing planets as a function of the scaled signal, both
including (solid lines) and omitting (dotted lines) a constant
acceleration term.  In each row of panels, we restrict our attention
to a small range of planetary orbital periods.  The far-left column of
panels is for detecting the planet, the center-left column is for
measuring the mass with 30\% accuracy, the center-right column is for
measuring the mass with 10\% accuracy, and the far-right column is for
measuring the orbital parameters with 10\% accuracy.  We have extended
the range of planetary orbital periods up to 10 years (twice the
survey duration) for Figs.\ \ref{FigRateVsS}-\ref{FigRateVsPConstM}.

For small values of the scaled signal, the planet is not detected,
regardless of the orbital period or the inclusion of an acceleration
term.  For sufficiently large values of the scaled signal, the planet
can be detected and characterized even for orbital periods somewhat
longer than the duration of the astrometric survey.  However, for
intermediate values of the scaled signal and periods comparable to
the mission lifetime, there can be a significant
reduction in the rates for detecting planets and measuring their
orbits, when an unknown constant acceleration is
included.

\subsection{Rates versus Planetary Orbital Period}

In Fig.\ \ref{FigRateVsPConstS} we plot the rates for detecting and
characterizing planets as a function of the planetary orbital period.
In each panel the solid line is for models including an unknown
constant acceleration and the dotted line assumes no acceleration.  In
each row of panels we fix the value of the scaled signal.  While
there is little difference for planetary orbital periods less than 2.5
years (half the survey duration), the presence of a wide binary
companion can dramatically lower the detection efficiency for longer
planetary orbital periods.  The narrow dips for small orbital periods
are aliasing due to observing at nearly periodic intervals.

Since the scaled signal is influenced by both the planet mass and
orbital period, we also consider the rates for detecting and
characterizing planets as a function of the planetary orbital period,
after dividing the population into subsets according to the mass of
the planet (see Fig.\ \ref{FigRateVsPConstM}) The rates for detecting
giant planets and measuring their masses and orbital parameters will
not be significantly affected by the presence of wide binary
companions (since we expect most giant planets to cause a large scaled
signal or be in a short-period orbit).  However, detectable planets
with masses less than $20M_\oplus$ are expected to induce a 
modest scaled signal and/or to have orbital periods comparable to the
duration of the astrometric survey.  Thus, the rates for detecting
terrestrial mass planets and measuring their orbits may be
significantly reduced when it is necessary to allow for the
acceleration of a wide binary companion.

\subsection{Rates versus Planetary Orbital Eccentricity}

\begin{figure}[ht]
\plotone{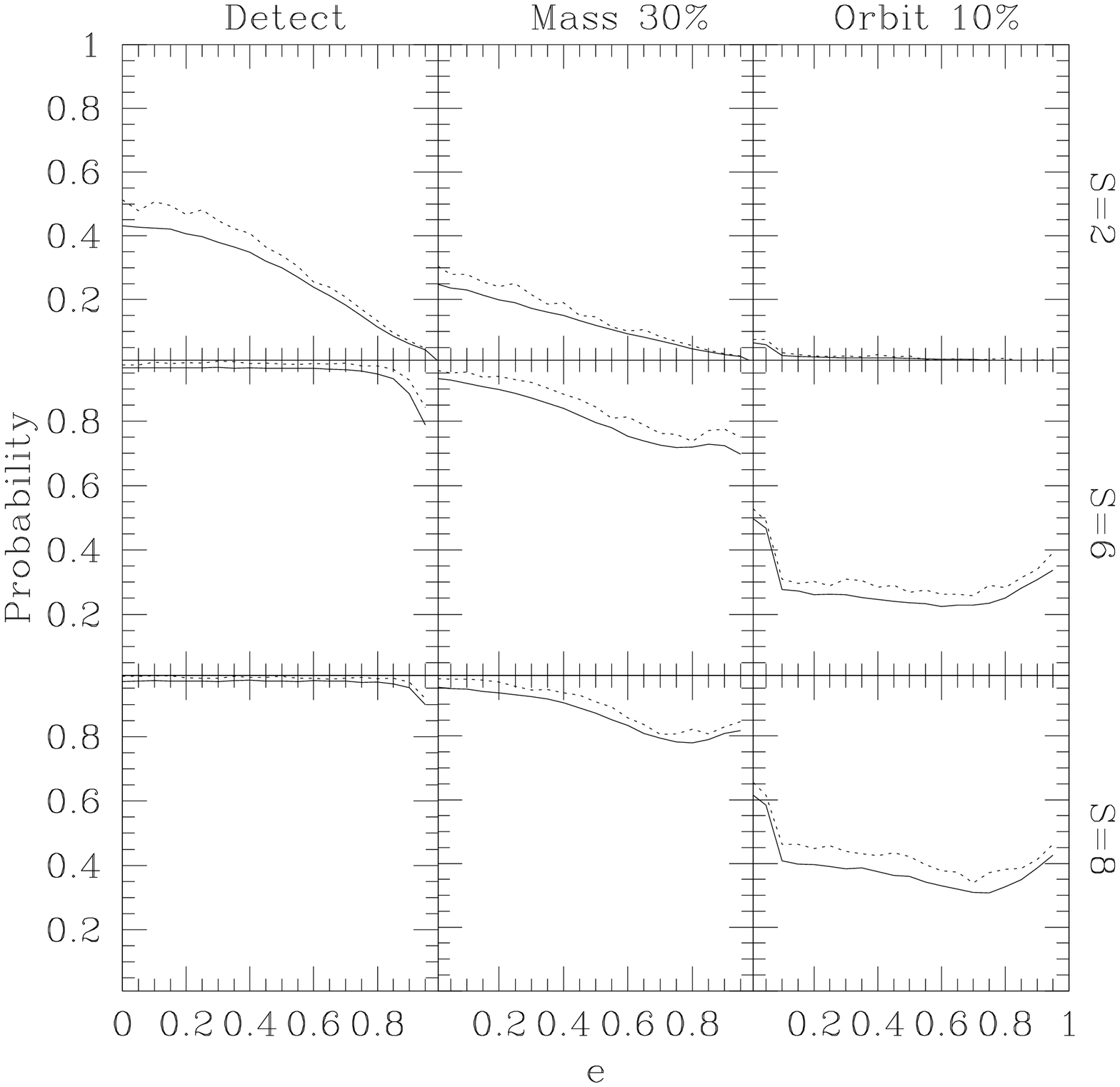} 
\caption[f5.eps]{
Here we show the rates for detecting and characterizing planets as a
function of the planetary orbital eccentricity.  In each panel the
solid line is for models including an unknown constant acceleration
and the dotted line assumes no acceleration.  Each row of panels is
for a different fixed value of the scaled signal (top: $S=2$, middle:
$S=6$, bottom: $S=8$).  The left column of panels is for detecting
the planet, the center column is for measuring the mass with 30\%
accuracy, and the right column is for measuring the orbital
parameters with 10\% accuracy.
\label{FigRateVsE}}
\end{figure}

Next, we investigate the effect of a binary companion on the rates for
detecting planets and measuring their masses and orbital parameters as a
function of the planetary orbital eccentricity (see Fig.\
\ref{FigRateVsE}).  While the rates do have a weak dependence on the
eccentricity, they are not significantly altered by allowing for an
unknown constant acceleration term.

\subsection{Rates versus Planetary Mass}

\begin{figure}[ht]
\plotone{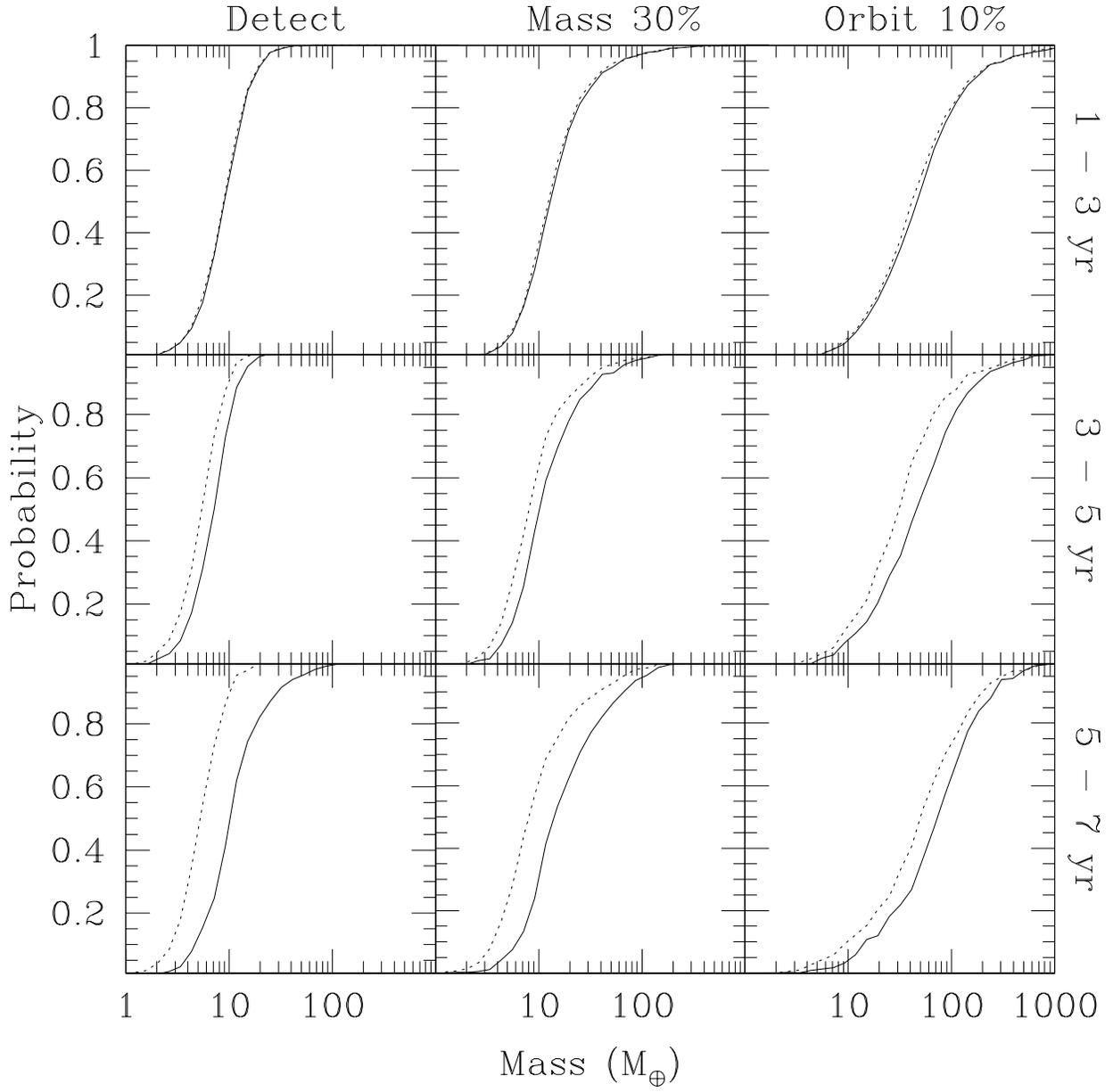}
\caption[f6.eps]{
Here we show the rates for detecting and characterizing planets as a
function of the planetary mass.  In each panel the solid line is for
models including an unknown constant acceleration and the dotted line
assumes no acceleration.  Each row of panels is for planets with
orbital periods in a given range (top: 1-3 yr, middle: 3-5 yr, bottom:
5-7 yr). The left column of panels is for detecting the planet, the
center column is for measuring the mass with 30\% accuracy, and the
right column is for measuring the orbital parameters with 10\%
accuracy.
\label{FigRateVsM}}
\end{figure}

Finally, we show the effect of a binary companion on the rates for
detecting planets and measuring their masses and orbital parameters as
a function of the planetary mass (see Fig.\ \ref{FigRateVsM}).  In
each row of panels we focus on planets with orbital periods in a given
range (top: 1-3 yr, middle: 3-5 yr, bottom: 5-7 yr).  While these
calculations assumed a distance of 10pc and 24 pairs of 1-d
observations with single measurement precision of $1\mas$, the
results can be scaled to other distances and measurement precisions.
For orbital periods much shorter than the duration of the astrometric
survey, there is little effect due to the unknown acceleration term.
However, the rates of detecting planets and measuring their orbital
parameters can be significantly affected for low mass planets when the
orbital period approaches or exceeds the duration of the survey.

\section{Results for Planets in Multiple Planet Systems}

In this section, we present the results of simulated observations of
many stars with two planets to determine our ability to detect
planets and measure their masses and orbital parameters.  To
understand the effects of multiple planets, we compare the results
with simulations including only a single planet.  

%
We tabulate some of our results in Table 1 (electronic version only).
The table includes a subset of our results particularly relevant to
one of the questions that we address in this section.  We list the
mass and orbital period of each planet (named A and B) in the first
four columns.  In columns 5 and 6 we list the probabilities of
detecting planet A or B, if it were the only planet in the system.  In
columns 7 and 8 we list the efficiencies of detecting planets A and B
in a two-planet system with parameters given in columns 1-4.  We
define the efficiency for detecting a planet to be the probability of
detecting the planet when in the two-planet system divided by the
probability of detecting the planet if it were the only planet in the
system.  In column 9 we list the probability of detecting both planets
simultaneously.

In Figure 7 we show the probability contours for detecting a single
planet as a function of its orbital period and the scaled signal,
averaging over the mass and other parameters.  The dotted lines are
for rejecting the best-fit no-planet model with 99.9\% probability,
while the solid lines also require that the reduction in $\chi^2$ due
to using the best-fit one-planet model is significant at the 99.9\%
level.  Note that this paper requires a planet to pass the
latter test to be considered detected and that this is a more strict
requirement than was used by Sozzetti \etal (2002) and Ford \&
Tremaine (2003).  
%
%
The sharp rise in
required signal near $P=5$ yr is due to the duration of the simulated
astrometric survey.

In Figure 8 we show probability contours for detecting planets in a
two-planet system as a function of their scaled signals, averaging
over their masses, orbital periods (up to 5 years), and other
parameters.  The solid lines are for detecting both planets, and the
dashed line and dotted-dashed line are for detecting either planet
(regardless of whether the other planet is detected).  For reference,
the dotted lines show the probability contours for detecting a single
planet, if the other planet were not present.  The bold contours are
for a 50\% probability of detection.  When one planet has a small
scaled signal ($S \le 1$), it has only a slight effect on the
efficiency for detecting the other planet.  However, when one planet
has a significant scaled signal ($S \ge 4$), then the scaled signal
necessary to detect the other planet is significantly increased (by a
factor $\sim2.5$).

Next, we consider several questions about how the planet finding
capabilities of SIM will be affected by the presence of two planets in
light of our simulations.  While our numerical simulations cover a
wide range of parameter space, it is difficult to visualize our
results over the full multi-dimensional parameter space.  Therefore,
the discussion of our results concentrates on the most important
factors, the planet masses and orbital periods.  We focus on the
effects of additional giant planets ($\sim1M_{J}$) and low mass
planets ($\le20M_{\oplus}$) with orbital periods in one of three
categories, short ($\sim3$d), intermediate ($\sim1-3$yr), or long
($\sim12$yr).

\subsection{Is a giant planet harder to detect when a long-period planet in present?}

First, we consider systems with one giant planet in a $1-3$yr orbit and an additional
planet with a 12 year orbital period.
The inner giant planet
is easily detectable in the absence of the outer planet, and
the efficiency is only slightly affected by the presence of a distant outer companion.

In Figure 9 we plot the probability of detecting a planet with a 1
year orbital period as a function of the mass of that planet, $m_1$.
The solid line is for a single planet, while the other lines are for a
two-planet system with a second planet of various masses in a 12 year
orbit.  The main effect of the outer planet is to raise the mass
of the smallest detectable inner planet by a factor of $\sim3$.

In Figure 10 we plot the probability of detecting an inner planet as a
function of the planet's orbital period, $P_1$.  The solid lines are
for the planet by itself, and the dotted lines are for the inner
planet when there is a second planet in a long-period orbit ($P_2 =
12$ yr).  The different rows of panels are for detecting a planet of
different masses, and the different columns of panels are for
different masses of the outer planet.  For an inner planet with mass
$1 M_J$ (top), the efficiency is only slightly reduced for orbital
periods less than the mission duration in the presence of an outer
planet.  (In the top panels, the solid line is very near 1 for all
periods.)


\subsection{Is a giant planet harder to detect when a short-period planet is present?}

Next, we consider systems with one giant planet in a $1-3$yr orbit and an additional
planet with a 3 day orbital period.
The outer planet is
easily detectable in the absence of the inner planet, but the
probability of detecting the outer planet is slightly reduced ($\sim10\%$)
due to the presence of the inner planet.  

In Figure 11 we plot the probability of detecting a planet in a 1 year
orbit as a function of the mass of a second planet, $m_2$ in a
short-period orbit, $P_2=3$ d.  The horizontal lines show the
probability of detecting the outer planet in the absence of the inner
planet for reference.  The solid and dotted lines (top) are for an
outer planet with mass $m_1 = 1 M_J$ and the short and long-dashed
lines (bottom) are for an outer planet of mass $m_1 = 20 M_\oplus$.
The probability of detecting the outer planet is somewhat reduced,
even for a $1 M_J$ mass outer planet in the presence of a low-mass
planet in a 3 day orbit.  The probability of detecting a $20 M_\oplus$
outer planet is dramatically reduced by the presence of a short-period
planet with mass $m_2 \ge 0.3 M_J$.  As we discuss in \S5, adding a
modest number of radial velocity observations can often restore the
sensitivity for detecting such planets.

In Figure 12 we plot the probability of detecting an outer planet as a
function of the planet's orbital period, $P_1$.  The solid lines are
for the planet by itself, and the dotted lines are for an outer planet
when there is a second planet in a short-period orbit ($P_2 = 3$ d).
The different rows of panels are for detecting a planet of different
masses, and the different columns of panels are for different masses
of the inner planet.  For an outer planet with mass $1 M_J$ (top), the
efficiency is only somewhat reduced ($\sim20\%$) when in the presence
of an inner planet.


\subsection{Is a giant planet harder to detect when another giant planet is nearby?}

We now consider systems with two giant planets with roughly
comparable orbital periods (e.g., 1 and 3 years).
These
planets would be easily detectable in isolation and are only slightly
affected by the presence of a nearby giant planet.  This is the type
of two-planet system for which it is most likely that SIM will be able
to detect both planets.

In Figure 13 we plot the probability of detecting a planet (1) in a 1
year orbit as a function of the mass of a second planet, $m_2$ in a 3
year orbit.  For reference, the horizontal lines are for the inner
planet in isolation.  The top pair of lines is for $m_1=1 M_J$, and
the bottom pair of lines is for $m_1=20 M_\oplus$.  A $1 M_J$ inner
planet (top) is only slightly affected, but a 20 $M_\oplus$ inner
planet (bottom) can become dramatically more difficult to detect when $m_2 \ge
10 M_\oplus$.  As we discuss in \S5, adding radial velocity observations can often restore the sensitivity for detecting such planets.

In Figure 14 we plot the probability of detecting a planet as a
function of its orbital period, $P_1$.  The solid lines are for a
single planet, while the dotted lines are for a two-planet system with
a second planet in a 3 year orbit.  The different rows of panels are
for detecting an inner planet of different masses and the different
columns are for different mass of the outer planet.  The probability
of detecting a $1 M_J$ planet (top left) can be somewhat affected for
some orbital periods, but it is only slightly affected for orbital
periods between $\sim0.5$ and $5$ years, where SIM will be most sensitive.


\subsection{Is a giant planet harder to detect when low-mass planet is nearby?}

It is similar for systems with one giant planet and one low-mass
planet with roughly comparable orbital periods (e.g., 1 and 3 years).
The giant planet is still easily detectable for most orbital
periods, and there is even a reasonable chance of detecting the lower
mass planet for some masses and orbital periods.  There is only a small
decrease in efficiency for orbital periods near the orbital period of
the low-mass planet (see Figure 13, top center and top right).  

\subsection{Is a low-mass planet harder to detect when a long-period planet is present?}

The probability of detecting a low-mass planet is significantly
reduced in the presence of a long-period outer planet.
For an inner planet with mass $20 M_\oplus$, there is a significant
reduction in the efficiency across the entire range of orbital periods
that SIM will probe (see Fig.\ 10, bottom).  In particular, a $1 M_J$
mass planet with a $12$ year orbital period drastically reduces the
probability of detecting a $20 M_\oplus$ mass planet with a $2-5$ year
orbital period from close to $100\%$ to $20-60\%$.  


\subsection{Is a low-mass planet harder to detect when a short-period planet is present?}

The probability of detecting a low-mass planet can also be reduced due
to the presence of a short-period inner planet.
The effect is somewhat significant ($\sim10\%$) for low-mass
short-period planets, and it becomes extremely difficult to detect an
outer planet with mass $20 M_\oplus$ for a sufficiently massive
short-period planet (see Fig.\ 11, short-dashed curve).  For example,
the presence of a $1 M_J$ inner planet reduces the probability of
detecting a $20 M_\oplus$ planet with an orbital period near 5 years
by $\sim25\%$ and by even more for shorter orbital periods (see Fig.\
2, bottom left).  On the other hand, the presence of a $20 M_\oplus$
planet in a short-period orbit, results in only a $\sim15\%$ reduction
in the probability of detecting a $20 M_\oplus$ planet with orbital
period $\sim1-5$yr (see Fig.\ 6, bottom right).  Fortunately, the
addition of radial velocity observations can significantly improve the
probability of detecting the low-mass planet, as discussed in \S5.


\subsection{Is a low-mass planet harder to detect when a giant planet is nearby?}

The probability of detecting a low-mass planet can also be reduced due
to the presence of a nearby giant planet.
For example, detecting a $20 M_\oplus$ planet in a 1 year orbital
period is very unlikely in the presence of a planet with mass $\ge6
M_\oplus$ with a 3 year orbital period (see Fig.\ 13, dashed curves).
This causes a decrease in efficiency by $\ge50\%$ over the entire
range of orbital periods that SIM will probe (see Fig.\ 14, bottom
left).  Again, the addition of radial velocity observations can
significantly improve the probability of detecting the low-mass planet
(see \S5).


\subsection{Is a low-mass planet harder to detect if another low-mass planet is nearby?}

The presence of two nearby low-mass planets also reduces the
probability of detecting either planet.
For example,
for two $20 M_\oplus$ planets with orbital periods of 1 and 3 year,
there is a $\ge50\%$ reduction in the probability to detect either
planet (see Fig.\ 14, bottom center).  For a pair of planets with
masses $20 M_\oplus$ and $1-5 M_\oplus$, the effect is smaller and
only occurs for a portion of the orbital periods that SIM will probe
(see Fig.\ 14, bottom right).


\section{Adding Radial Velocity Observations}

Combining an astrometric planet search with radial velocity
observations can result in an increased sensitivity to low mass planets
(Eisner \& Kulkarni 2002).  Just as we defined $S$ as the scaled
signal for astrometric observations, we can define
\be
R = \frac{m}{M} \frac{2\pi a}{P} \frac{\sin i}{\sigma_{RV}},
\ee
where $\sigma_{RV}$ is the single measurement precision for a radial
velocity observation.  When $R\gg S$, radial velocity observations are
typically more efficient at detecting a planet and measuring the five
orbital parameters accessible from radial velocities alone (except for
a small fraction of orbits with pathological orientations).  When
$S\gg R$, astrometric observations are more efficient for detecting a
planet and measuring it's mass and orbital parameters.  The transition
between these two regimes is independent of the star and planet masses
and occurs at an orbital period
\be
P_t = 0.30 \mathrm{yr} \sin i \left(\frac{D}{10\mathrm{pc}}\right) \left(\frac{\sigma_{SIM}}{1\mas}\right) \left(\frac{\mathrm{m/s}}{\sigma_{RV}}\right).
\ee
Eisner \& Kulkarni (2002) found that combining astrometric and radial
velocity observations for a single planet near this transition region
results in an improvement in sensitivity comparable to that which is
expected by simple scaling laws.  A simplistic analysis might suggest that
combining astrometric and radial velocity observations will provide a
significant improvement only for a relatively small fraction of
planets with orbital periods near this transition region ($0.5 P_t \le
P \le 2 P_t$), and that radial velocity
observations are of relatively little value for searching the
habitable zones of nearby stars (Shao \etal 2002).  However, this
argument does not apply to multiple planet systems.

Radial velocity surveys have shown that a star with one giant planet
within $\sim3$AU is more likely to have a second giant planet within
$\sim3$AU than a randomly chosen star of similar metallicity and
spectral type (Fischer \etal 2001).  Based on the currently known radial velocity planets,
at least $\sim13\%$ of planetary systems have multiple planets and
$\sim26\%$ of known planets are in a multiple planet system.  Since
additional planets may be found orbiting stars with a only single
currently known planet, these fractions are likely to increase as
additional planets are discovered by either radial velocity or
astrometric planet searches.  For planets in multiple planet systems,
the combination of astrometric and radial velocity observations could
be particularly valuable, as one planet may be easier to detect with
radial velocity observations and another planet may be easier to
detect with astrometric observations.  For example, many of the
currently known multiple planet systems have one giant planet in a
short-period orbit and a second more distant giant planet.  For such
systems, radial velocity measurements can easily constrain five of the
seven orbital parameters for the inner giant planet, allowing the
astrometric observations to more easily detect a more distant planet.

We have performed additional simulations to quantify the improvement
obtained from adding radial velocity observations
to a SIM planet search.  Our methods closely follow those described in
\S2, but we add 12, 24, or 48 radial velocity observations at random
times during the astrometric survey.  We also include a Gaussian noise
source with standard deviation, $\sigma_{RV} =3$m/s, the single
measurement precision for radial velocity observations.  Current
radial velocity searches have demonstrated a measurement precision of
$\simeq1$m/s, but stellar rotation and activity typically set the
effective precision for measuring the gravitational perturbation due
to planetary companions near $\simeq1-5$m/s, even for inactive stars (Wright 2005).  The
orbital fitting is done as before, fitting the orbits of both planets
to the astrometric and radial velocity observations simultaneously.
As for our simulations of purely astrometric surveys, we require that
the reduction in $\chi^2$ be significant according to an $F$-test for
a planet to be detected.  Note that since we are interested in systems
with multiple planets, methods based on the areal constant are not
applicable (Pourbaix 2002).

First, we present our results for systems with one planet in a 1 year
(left) or 3 year (right) orbit and an additional giant planet in a
short-period orbit (3 days).  In Fig.\ 15 we show the probability of
detecting the more distant planet as a function of its mass
(regardless of whether the short-period planet is detected).  The
inner planet has a mass of $1M_J$ (top) or $20M_\oplus$ (bottom).  The
solid line style shows the probability of detecting the more distant
planet for a survey with no radial velocity observations, while the
other line styles show the improvement from adding 12 (dotted), 24
(dotted-dashed), or 48 (dashed) radial velocity observations.  Without
the radial velocity observations, the presence of a short-period giant
planet (top panels) would significantly reduce SIM's sensitivity to
additional low-mass planets.  However, even a modest number of radial
velocity observations can fully compensate for the additional
complexity of a two-planet model.  For a $20M_\oplus$ short-period
planet (bottom panels), the effects are less significant.

Next, we present our results for such two-planet systems as a function
of the mass of the short-period planet (See Fig.\ 16).  We show
results for simulations where the more distant planet has a mass of
$0.3M_J$ (top panels), $20M_\oplus$ (upper middle panels),
$10M_\oplus$ (lower middle panels), and $5M_\oplus$ (bottom panels).
The columns and line styles are the same as for Fig.\ 15.  When the
short-period planet has a sufficiently small mass, even radial
velocity observations have difficulty constraining its orbit, so the
radial velocity observations can provide only a modest improvement in
the probability of detecting the more distant planet.  When the
short-period planet is sufficiently massive, adding radial velocity
observations can dramatically improve the probability of detecting a
more distant planet.  This effect is particularly impressive for
relatively low-mass planets ($\sim5-50M_\oplus$ at 10pc) which are a
prime target for the SIM planet searches.  The dip in the curves at
intermediate masses occurs when the perturbations from the two planets
are comparable, making it more difficult to separate the two signals.


\section{Discussion}

\subsection{Wide Binary Stars}

When we allow the no-planet model to include an unknown constant
acceleration term appropriate for a wide binary companion, the
sensitivity of our simulated astrometric surveys is unaffected for
planets with orbital periods much shorter than the duration of the
survey, but the sensitivity is significantly decreased for orbital
periods approaching the duration of the survey.  For planets with
orbital periods longer than the duration of the survey, the reduced
sensitivity due to an unknown acceleration can be much larger.  While
the unknown acceleration due to a distant binary companion will
decrease the rate at which very low mass planets are detected, the
efficiency for detecting giant planets around nearby stars will be
only slightly reduced, since these either have a short orbital period
(small semi-major axes) or a large signal (large semi-major axis).

These results have important consequences for choosing targets for an
astrometric planet search such as those planned for SIM.  If the
primary purpose of a $5$ yr survey were to detect low mass planets
with orbital periods less than $3$ yr, then it would be advantageous
to target nearby wide binary stars rather than more distant single
stars.  This would  be particularly relevant if searching for planets in the
habitable zone of a solar-type star.  On the other hand, if searching
for low mass planets with orbital periods slightly less than the
survey duration, then the presence of a wide binary companion is
likely to significantly reduce the sensitivity.  This could be
particularly relevant if searching for a terrestrial mass planet near
the habitable zone of an A or F star, as suggested by Gould, Ford, \&
Fischer (2003).  If the SIM mission were extended to ten years, then
the unknown acceleration due to a wide binary companion would not
cause a significant reduction in sensitivity when searching for
planets in the habitable zones of nearby stars, regardless of spectral type.

\subsection{Two-Planet Systems}

When planning the SIM planet searches, it will also be important to
consider the effects of multiple planet systems.  We have presented
results of simulated observations for a wide variety of two-planet
systems.  
First, we make some limited comparisons to the results of Sozzetti
\etal (2003).  Sozzetti \etal (2003) claim that the sensitivity for
multiple planet systems with scaled signal $S\sim1$ is not
significantly reduced compared to single planet systems, but they
tolerate an increased false alarm rate (of up to 10\%).  We hold the
false alarm rate fixed (at 0.1\%), and find that then the scaled
signal required to detect a planet increases when there are multiple
planets.  While this makes detailed comparisons difficult, it
illustrates an important capability of the SIM mission.  While solid
detections (e.g., low false alarm probability, $\sim 0.1\%$) are an
important goal of the SIM planet searches, there is also significant
value in the inevitable marginal detections (e.g., false alarm
probability, $\sim50\%$).  These marginal detections can provide an
enriched sample of target stars for future direct imaging planet
searches, such as TPF-C and TPF-I (Marcy \etal 2005).  Additionally,
tentative estimates of orbital parameters could be useful for
maximizing the value of such observations.  For example, many planets
in the habitable zone will only be visible for a modest fraction of
their orbit when they are near maximum separation from their star, as
viewed from the Earth.  Being able to time observations so as to avoid
times when a planet is too close to the star for direct imaging, would
save valuable observing time, allowing a greater number of stars to be
searched more thoroughly.  Similarly, preliminary estimates of a
planet's orbital geometry could be used to choose telescope
orientations which maximize their chance of imaging the planet and
eliminate the need to take multiple images at various roll angles.
Such synergies between SIM and direct imaging planet searches are an
intuitive example of the value of the more general and rigorous
framework provided by Bayesian adaptive experimental design that has
previously been described in the context of radial velocity surveys
(Ford 2005b) and purely astrometric surveys (Loredo 2004).  We leave a
more thorough investigation of the value of marginal detections and
synergies between various detection techniques for future study.
%

Our simulations show that the probability that SIM will detect a
planet is typically reduced when a second planet is present.  For
giant planets with orbital periods which SIM will probe, this is only
a small effect for most masses and orbital periods (up to
the mission duration) of the second planet.
The probability of SIM detecting a low-mass planet is likely to be
more significantly affected by an additional planet.  While there is a
high probability that SIM will detect single planets with masses
$\le20 M_\oplus$ and orbital periods $0.5\le P\le 5$ years, the
probability can be significantly reduced by a second planet in a wide
variety of locations.  A giant second planet can significantly reduce
the probability of detecting a low-mass planet, regardless of whether
the giant planet has an orbital period of 3 days, 1 year, or 12 years.
Fortunately, adding a radial velocity observations
can dramatically improve SIM's sensitivity to low-mass planets in
multiple planet systems.

For a purely astrometric survey, a no-planet model can be rejected
50\% of the time when any planet causes a scaled signal of $S\ge2.2$
(Sozzetti \etal 2001; Ford \& Tremaine 2003).  When a single planet
causes a scaled signal $S\ge4.6$, the masses and orbital parameters
are measured sufficiently accurately that the best-fit one planet
model significantly reduces $\chi^2$ (compared to the best-fit
no-planet model) 50\% of the time (Ford \& Tremaine 2003).  However,
if there is a second planet in the system, then a one-planet model
often does not result in a significant improvement in $\chi^2$ unless
one planet causes a scaled signal of $S\ge12$.  When the planet
inducing the smaller perturbation causes a scaled signal $S_<\le4$, it
may be possible to detect the planet inducing the larger scaled signal
for $4.6\le S_>\le12$.  To obtain a two-planet model which
significantly reduces $\chi^2$ (compared to the best-fit one-planet
model) 50\% of the time, both planets need to cause a scaled signal
$S\ge12$ (See Fig.\ 2).  When a radial velocity
observations are added, the sensitivity of the
combined survey becomes a function of the scaled signals and the
orbital periods of both planets, making it difficult to summarize the
results in a single figure.  We have included figures that
illustrate the sensitivity of combined surveys for several cases of
particular interest (see Figs.\ 15 \& 16).  We find that such surveys
are much more sensitive to low-mass planets (by a factor $\sim2$--3
in mass) when there is an additional giant planet, particularly when
the giant planet is in a short-period orbit.  Thus, the combined
astrometric and radial velocity survey typically has a sensitivity to
low-mass planets in two-planet systems that is comparable to
sensitivity to single low-mass planets for the astrometric survey
alone.

\subsection{Hierarchical and Many-Planet Systems}

In this paper we have simulated systems with just three massive bodies
(one star and two planets or two stars and one planet), and the large
parameter space already makes a thorough study impractical.
Therefore, we have not attempted to carry out simulations for four or
more massive bodies, despite the fact that a few such planetary
systems have already been discovered.  Nevertheless, we can make some
general comments about the challenges that such systems will create.
Searching for planets around one star in a hierarchical triple (or
higher multiple) star system should be very similar to searching for
planets around one star in a wide binary system.  Long term stability
of the planet requires that all stellar companions must be much
farther away, so all such stars will move only slightly over the
duration of observations.  Therefore, the perturbation of the
additional stars can be modeled as a single constant acceleration, and
our simulations of wide binary stars should be directly applicable..

For planetary systems with two planets and a wide binary companion, we
expect that the unknown acceleration due to the distant stellar
companion will only be significant when at least one planets has an
orbital period greater than half the mission duration.  For such
systems, the unknown acceleration will make it more difficult to
detect and characterize the orbit of the outer planet.  If the outer
planet induces a perturbation comparable to or larger than the
perturbation due to the inner planet, then it will also be more
difficult to detect and characterize the orbit of the inner planet.
In cases where the outer planet induces a perturbation small relative
to that of the inner planet, the detection and characterization of the
inner planet's orbit should be relatively robust to the unknown
acceleration.

For planetary systems with three or more planets, the situation is
more complicated.  A comprehensive study of all plausible
configurations is not feasible due to the enormous range of parameter
space.  Even a question-based approach similar to \S 4 would require
an unreasonably lengthy text.  Here we only raise a few qualitative
concerns.  Based on our experience analyzing radial velocity and
simulated astrometric data sets, it is essential for the number of
observations to be at least several times the number of free model
parameters.  For example, the radial velocity of a star orbited by
three planets on Keplerian orbits is described 16 model parameters,
yet the discovery papers announcing three planet systems were based on
a much larger number of observations, 137 observations for $\upsilon$
And, 146 for 55 Cnc, 76 for HD 160691, 52 for HD 37124 (for which two
qualitatively different orbital solutions were offered), and 155 for
GJ 876.  To model the astrometric motion of a star orbited by three
planets requires 26 model parameters (2 position, 1 parallax, 2
velocity, 3 masses, and 3 sets of 6 Keplerian elements).  Therefore,
we expect that it will be extremely difficult to detect and accurately
measure orbital parameters for three planet systems using only the
currently planned SIM observations of 50-100 one dimensional
observations per star over a five year mission.  Allocating additional
follow up observations during the first five years and/or extending the
mission lifetime to ten years would be extremely valuable for
characterizing such multiple planet systems.  For multiple planet
systems where at least one planet has an orbital period less than
$P_t$, radial velocity observations could significantly improve the
ability of SIM to detect and measure the orbital parameters of
additional planets at larger distances, by accurately measuring the
orbits of the short-period planets.

\subsection{Consequences for SIM}

Our results have significant consequences for the SIM planet searches.
On the positive side, SIM will often be able to detect and measure the
astrometric parameters of both planets in systems with two giant
planets with orbital periods $60$ d$\le P\le5$ yr.  However, if most
low-mass planets are typically in systems with multiple planets, then
SIM along would have more difficulty detecting the low-mass planets in
these systems without accompanying radial velocity observations.  As
discussed in \S5, the addition of a modest number of high precision
radial velocity observations can significantly increase the survey's
sensitivity for various types of planetary systems.  
This implies that it is very important that the SIM
planet search be accompanied by radial velocity observations whenever
possible (e.g., even for M stars that require long exposure times for
precise radial velocities).  While such observations are already
underway for many SIM planet search target stars, for other stars
(e.g., F \& A stars) it is more difficult to obtain precise radial
velocities (e.g., due to increased activity and rotational broadening
of spectral lines).  When interpreting the results of the SIM planet
searches, it will be important to consider the variation in SIM's
sensitivity to low-mass planets in multiple planet systems as a
function of the quality of radial velocity constraints.

If the mission lifetime of SIM were extended to 10 years, then the
sensitivity to low-mass planets is further improved.  However, the
improvement due to adding radial velocity information is reduced (see
Figs.\ 15 \& 16, red curves for 10 year surveys in electronic version
only).  Even if the baseline mission lifetime were extended to 10
years, we believe it would still be important to obtain at least a
modest number of radial velocity observations ($\sim12-24$), so that
low-mass planets can be identified early during the SIM mission,
permitting additional astrometric measurements to be obtained for
these particularly interesting system.  It is important to remember
that the probabilities presented in this paper are for detection only.
Additional observations will be necessary to precisely determine the
various orbital parameters, which are of particular interest for
multiple planet systems.  If such systems can be identified early in
the SIM mission, then it will be possible to allocate additional
observations that can significantly improve the measurement of
orbital parameters (e.g., Loredo 2004; Ford 2005b).

\subsection{Review of Key Assumptions}

It is important to keep in mind the assumptions and limitations of our
simulations.  The tables and figures presented assume a star 10pc away
is observed with 24 pairs of 1-d astrometric observations with single
measurement precision $\sigma_{SIM} = 1\mas$, as suggested by the SIM
planet search proposals (Marcy \etal 2002; Shao \etal 2002).  While
changing any of these parameters would affect the masses of planets which SIM
is likely to detect, we believe our simulations provide valuable
insight into how the presence of additional companions may affect
SIM's sensitivity.  For example, 88 of the stars
selected as SIM narrow angle planet target are closer than 10pc,
and 47 are closer than 6pc.  For these targets, our results showing
the sensitivity of SIM as a function of masses can be easily
rescaled for the appropriate distances.  Similarly, should the final
performance of SIM differ from the specifications, our results can be
rescaled for alternative values of the single measurement precision.
In one sense our simulations are optimistic, since we assume that good
initial estimates of the astrometric and orbital parameters are
available for a local search algorithm.  In practice, it will be
necessary to generate initial guesses for these parameters from the
observations themselves.  We believe that it is important for future
studies to develop such algorithms, particularly for multiple planet
systems.  One reason our simulations might underestimate the
sensitivity of SIM is due to the assumption of a non-adaptive
scheduling algorithm.  Early studies of adaptive scheduling algorithms
for radial velocity searches suggest that adaptive scheduling could
increase their sensitivity by a factor $\sim2$ (Ford 2005b; Loredo
2004).  While we believe it that such adaptive scheduling algorithms
will be essential to maximize the value of SIM observing time,
evaluating such algorithms is well beyond the scope of this study.

\acknowledgments

We thank Scott Tremaine for his guidance and suggestions for the
manuscript.  We acknowledge valuable discussions with Debra Fischer,
Geoff Marcy, and Chris McCarthy.
This research used computational facilities supported by NSF grant
AST-0216105, and was supported in part by NASA grant NAG5-10456, the
EPIcS SIM Key Project, and the Miller Institute for Basic Research.

\clearpage

\begin{figure}[ht]
\plotone{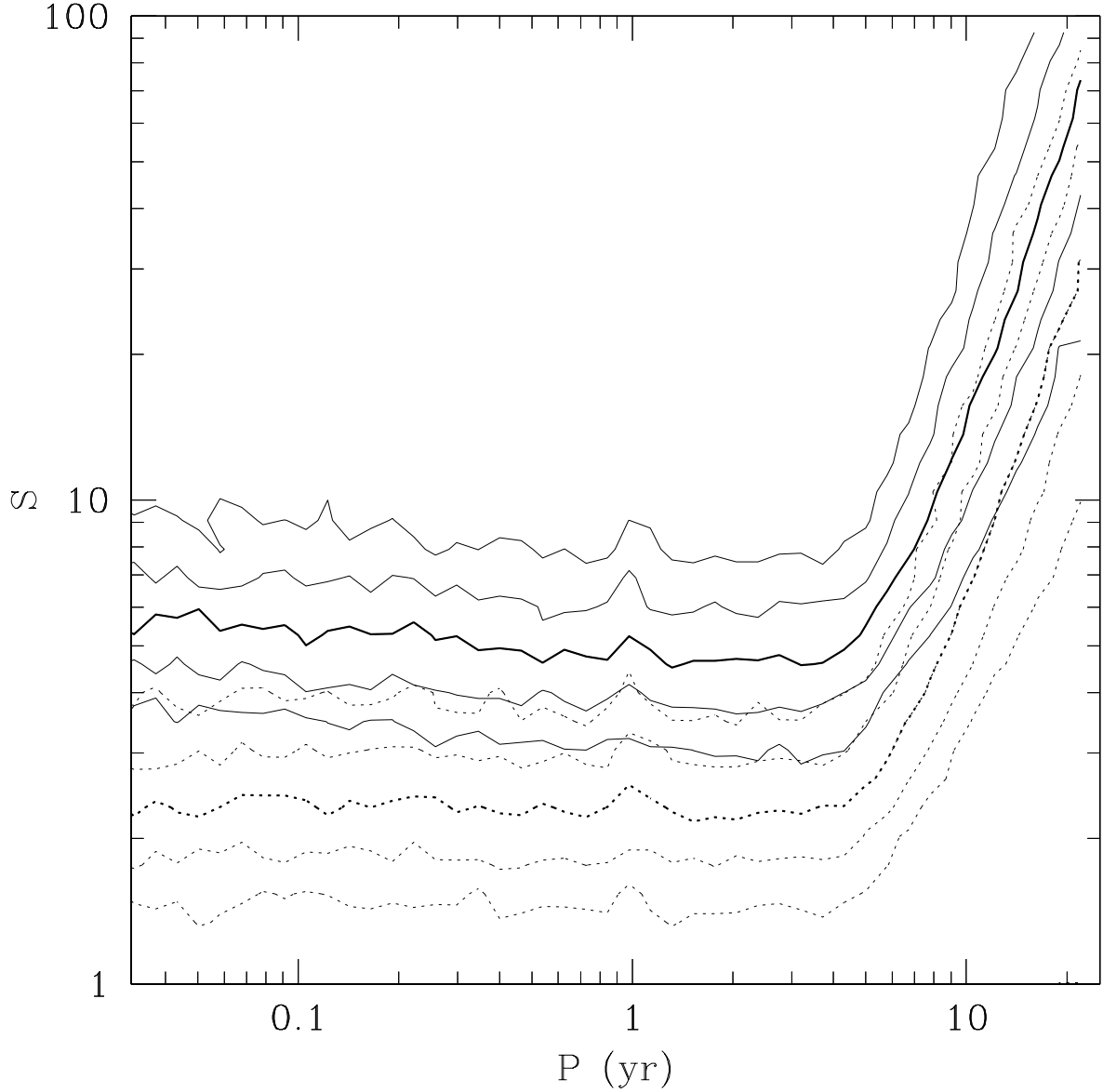} 
\caption[f7.eps]{
Here we show probability contours for detecting a single planet as a
function of its orbital period and the scaled signal, averaging over
the mass and other parameters.  The contours are for 10\%, 25\%, 50\%
(bold), 75\%, and 90\% probability (bottom to top).  The dotted lines
are for rejecting the best-fit no-planet model with 99.9\%
probability.  The solid lines also require that the reduction in
$\chi^2$ due to using the best-fit one-planet model is significant at
the 99.9\% level.  These probability contours are similar to the
probability contours for measuring the mass and orbital parameters
with 30\% accuracy 90\% of the time (not shown).
\label{Fig4a}}
\end{figure}

\begin{figure}[ht]
\plotone{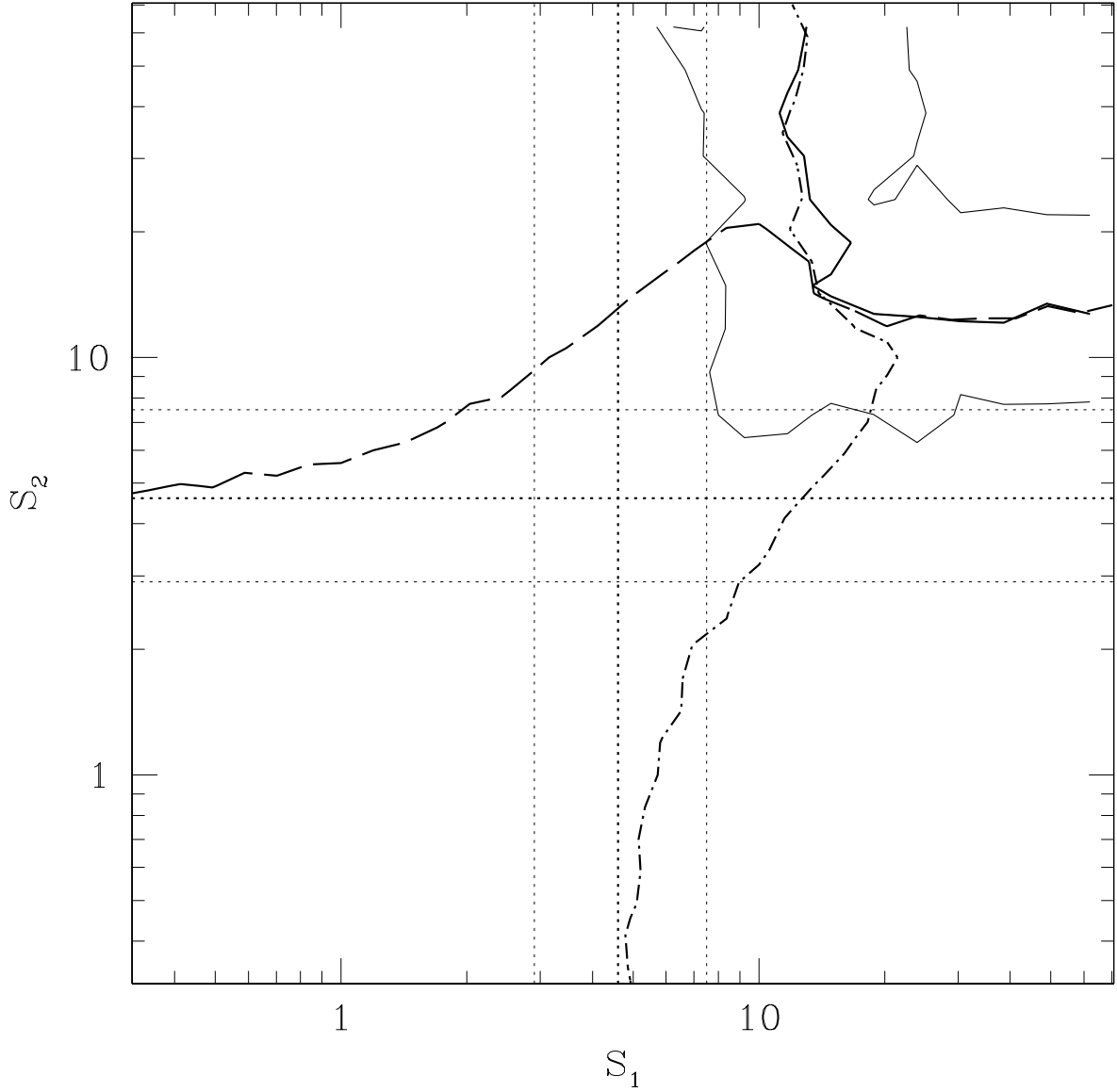} 
\caption[f8.eps]{
Here we show probability contours for detecting planets in a
two-planet system as a function of their scaled signals, averaging
over their masses, orbital periods (up to 5 years), and other
parameters.  The solid lines are for detecting both planets, and the
dashed line and dotted-dashed line are for detecting either planet (regardless of whether
the other planet is detected).  The dotted lines are probability
contours for detecting a single planet, if the other planet were not
present.  The contours are for 10\%, 50\% (bold), and 90\%
probability.  Only the 50\% contour is shown for detecting either
planet individually.
\label{Fig4b}}
\end{figure}

\begin{figure}[ht]
\plotone{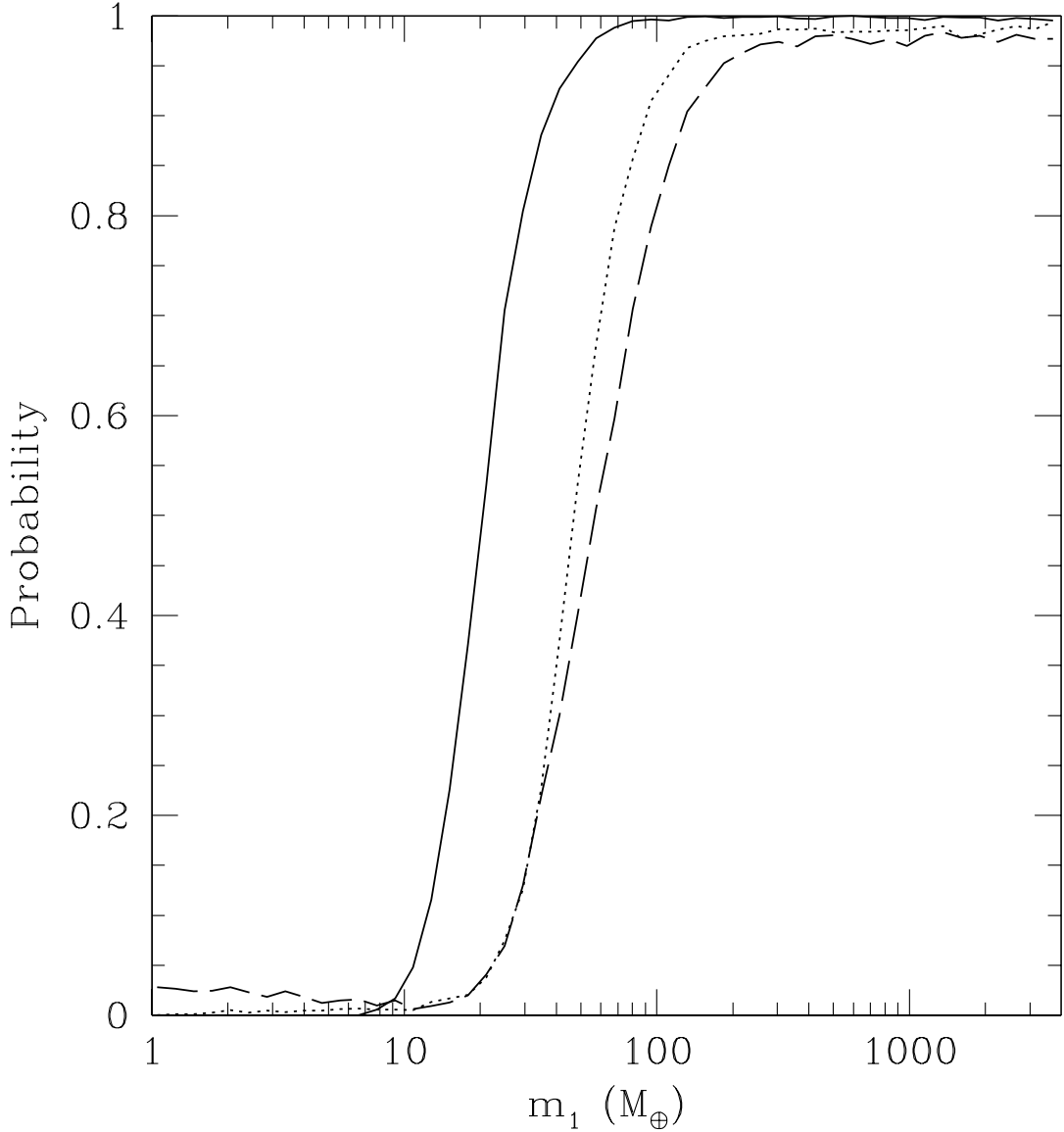} 
\caption[f9.eps]{
Here we show how the probability of detecting a planet is affected by
the presence of a second planet in a long-period orbit.  We plot the
probability of detecting a planet (1) in a 1 year orbit as a function
of the mass of that planet, $m_1$.  The solid line is for a single
planet, while the other lines are for a two planet system with a
second planet in a 12 year orbit (dotted line: $m_2=1 M_J$, dashed
line: $m_2=20 M_\oplus$).
Note that these calculations assumed a distance of 10pc
and 24 pairs of 1-d observations with single measurement precision of
$1\mas$, but the results can be easily scaled to other distances and
measurement precisions.

\label{Fig2a}}
\end{figure}

\begin{figure}[ht]
\plotone{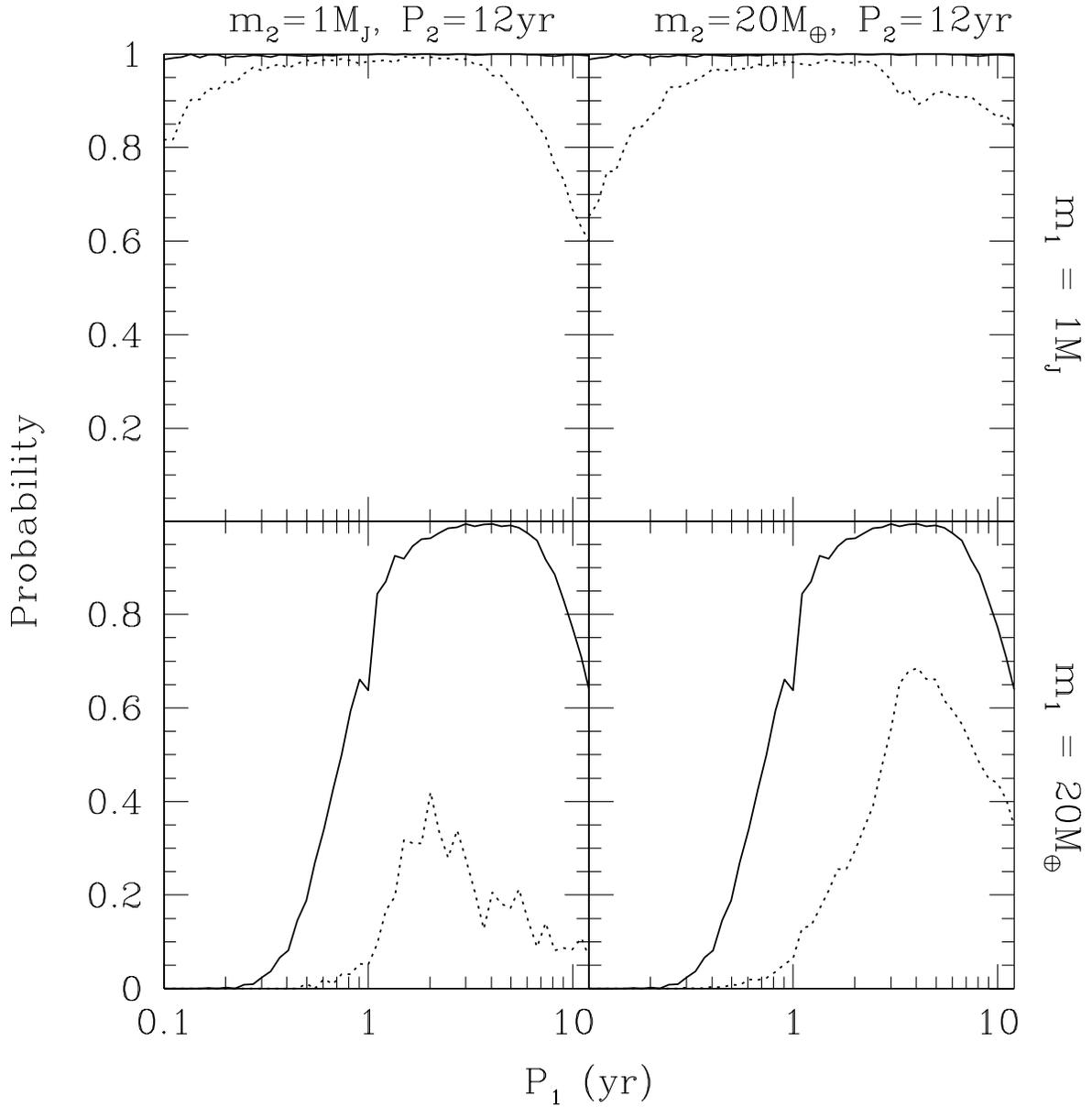} 
\caption[f10.eps]{
Here we show how the probability of detecting a planet is affected by
the presence of a second planet in a long-period orbit.
We plot the probability of detecting a planet (1) as a function of that planet's orbital period, $P_1$.
The different rows of panels are for detecting a planet of different
masses (top row: $m_1=1M_J$, bottom row: $m_1=20M_\oplus$).
The solid lines are for a single planet, and the dotted lines are for a two planet system.  
The left column of panels is for a second planet with mass $m_2=1 M_J$, 
and the right column of panels is for a second planet with mass $m_2=20 M_\oplus$.
\label{Fig2b}}
\end{figure}

\begin{figure}[ht]
\plotone{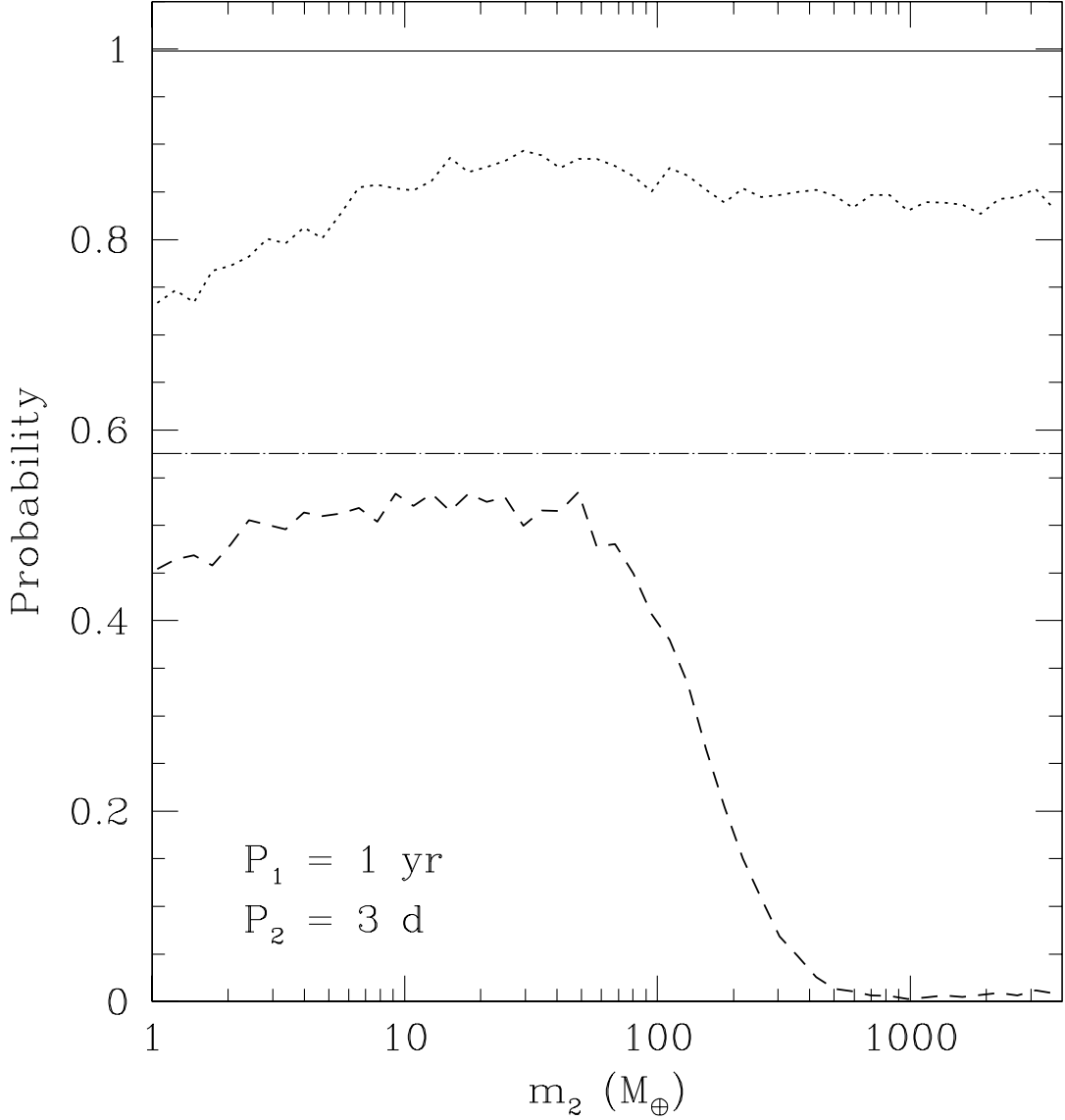} 
\caption[f11.eps]{
Here we show how the probability of detecting a planet is affected by
the presence of a second planet in a short-period orbit.  We plot the
probability of detecting a planet (1) in a 1 year orbit as a function
of the mass of the short-period planet, $m_2$.  The solid and dotted-dashed lines are
for the outer planet in isolation, while the other lines are for a two
planet system with a second planet in a 3 day orbit.  The top pair of
lines is for $m_1=1 M_J$, and the bottom pair of lines is for $m_1=20
M_\oplus$.
\label{Fig1a}}
\end{figure}

\begin{figure}[ht]
\plotone{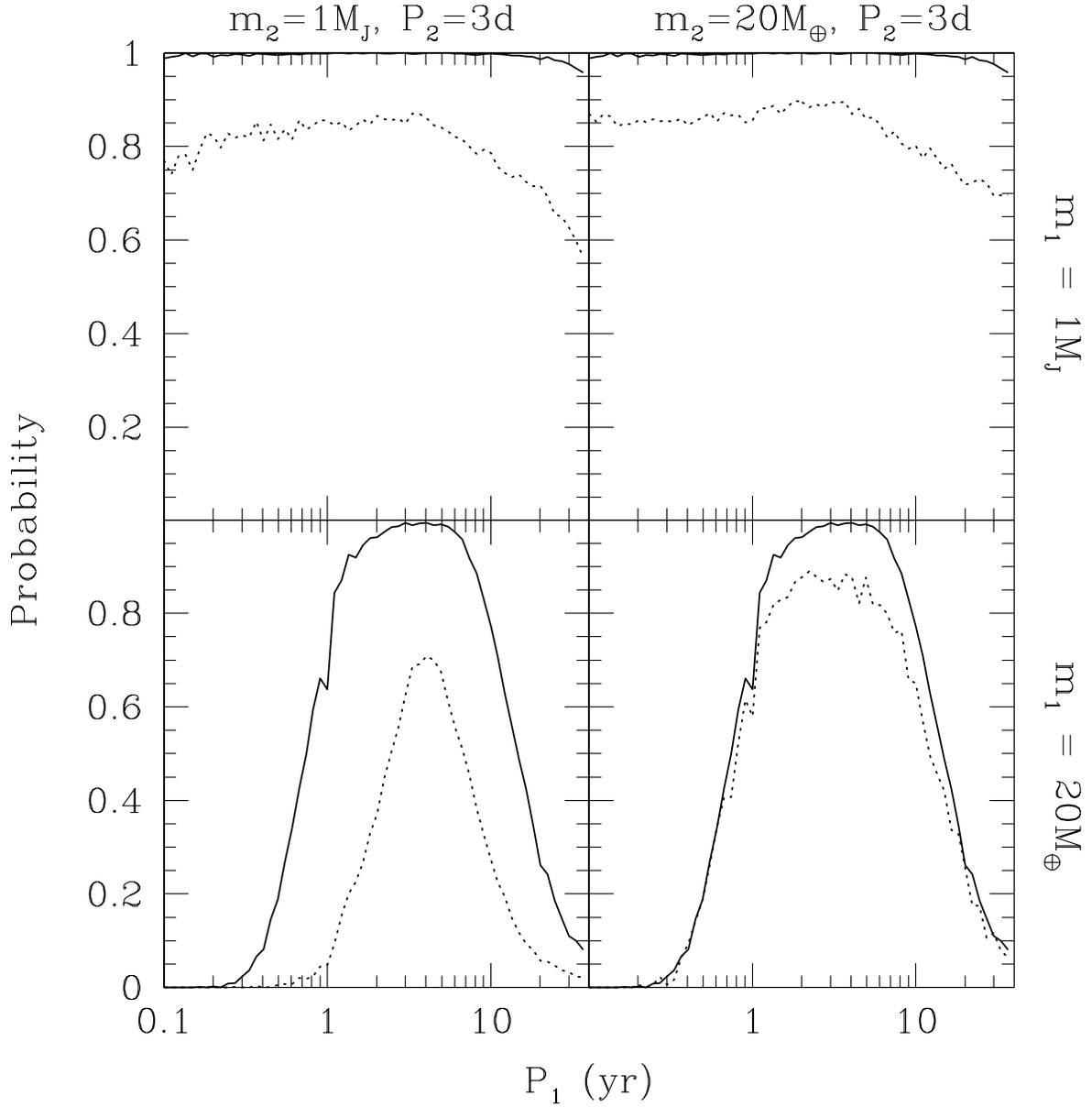} 
\caption[f12.eps]{
Here we show how the probability of detecting a planet is affected by
the presence of a second planet in a short-period orbit.  We plot the
probability of detecting a planet (1) as a function of the orbital
period of that planet, $P_1$.  
The different rows of panels are for detecting a planet of different
masses (top row: $m_1=1M_J$, bottom row: $m_1=20M_\oplus$).
The solid lines are for a single planet, while the dotted lines are
for a two-planet system with a second planet in a 3 day orbit.
The left column of panels is for a second planet with mass $m_2=1 M_J$, 
and the right column of panels is for a second planet with mass $m_2=20 M_\oplus$.
\label{Fig1b}}
\end{figure}

\begin{figure}[ht]
\plotone{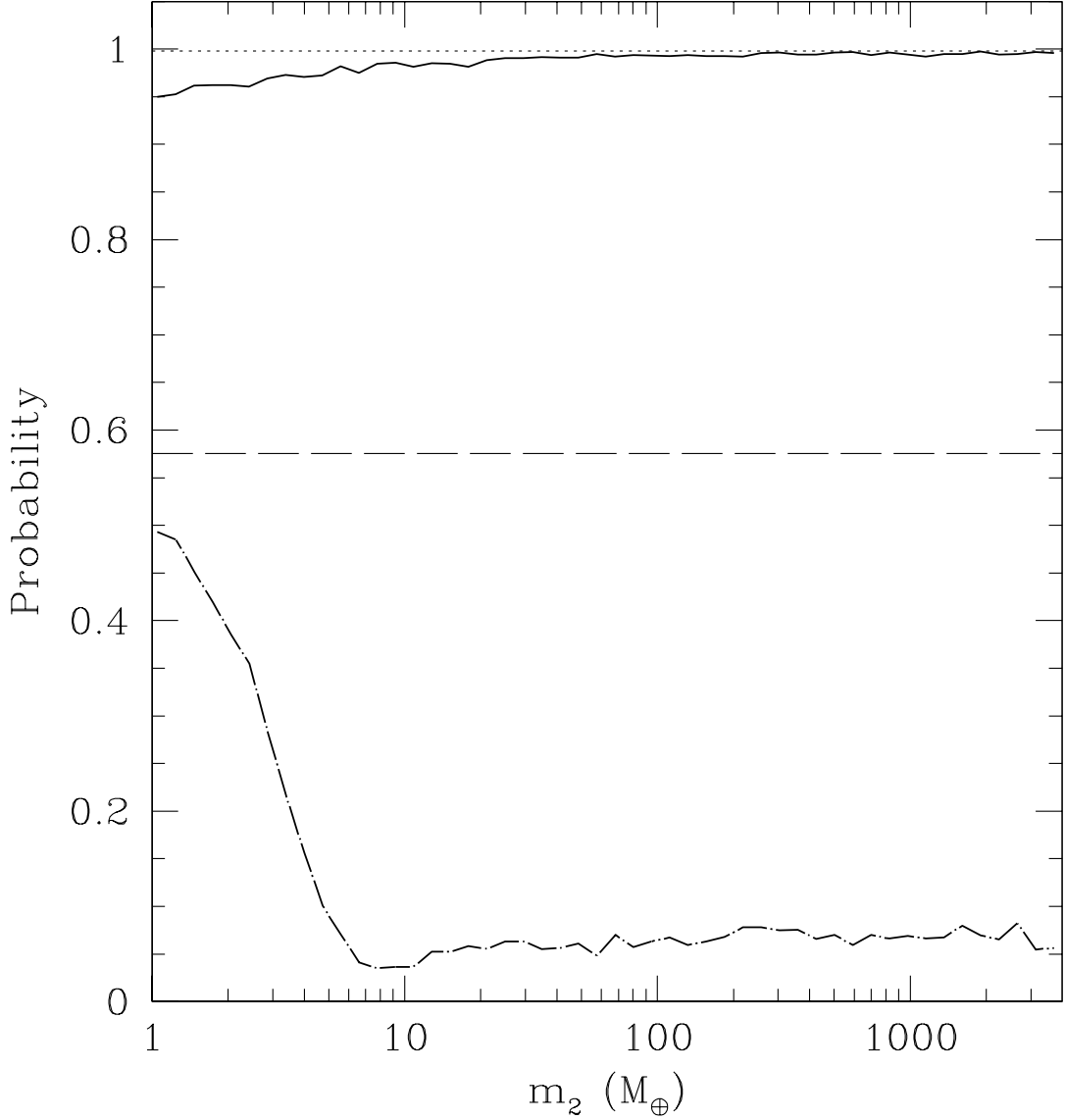} 
\caption[f13.eps]{
Here we show how the probability of detecting a planet is affected by
the presence of a second nearby planet.  We plot the probability of
detecting a planet (1) in a 1 year orbit as a function of the mass of
a second planet, $m_2$ in a 3 year orbit.  For reference, the
horizontal lines show the probability for detecting the inner planet
in isolation.  The other lines are for a two-planet system with a
second planet in a 3 year orbit.  The top pair of lines is for $m_1=1
M_J$, and the bottom pair of lines is for $m_1=20 M_\oplus$.
\label{Fig3a}}
\end{figure}

\begin{figure}[ht]
\plotone{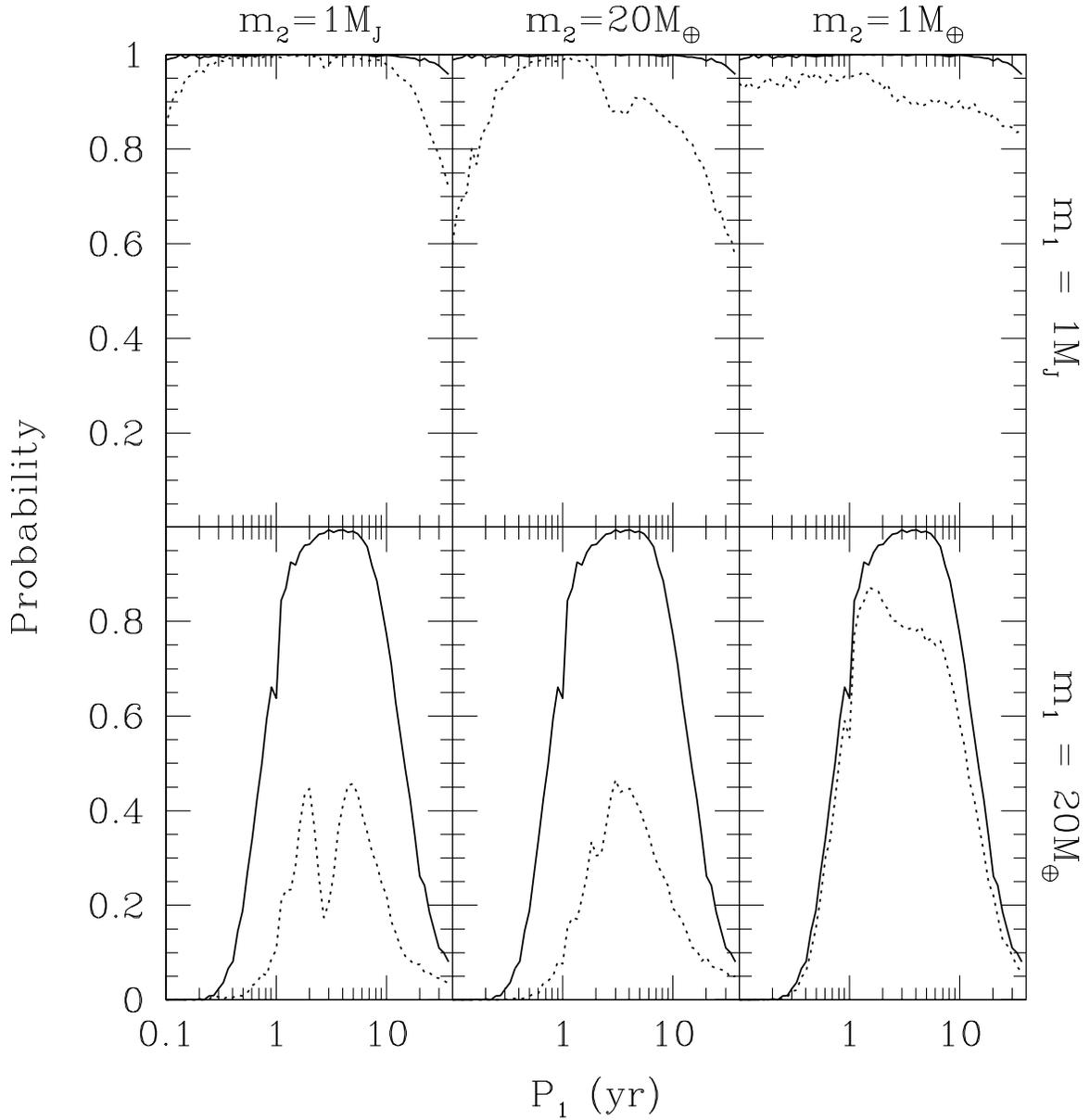} 
\caption[f14.eps]{
Here we show how the probability of detecting a planet is affected by
the presence of a second nearby planet.  We plot the
probability of detecting a planet (1) as a function of the orbital
period of that planet, $P_1$.  
The different rows of panels are for detecting a planet of different
masses (top row: $m_1=1M_J$, bottom row: $m_1=20M_\oplus$).
The solid lines are for a single planet, while the dotted lines are
for a two planet system with a second planet in a 3 year orbit.
The left column of panels is for a second planet with mass $m_2=1 M_J$, 
the center column of panels is for a second planet with mass $m_2=20 M_\oplus$,
and the right column of panels is for a second planet with mass $m_2=1 M_\oplus$.
\label{Fig3b}}
\end{figure}

\begin{figure}[ht]
\plotone{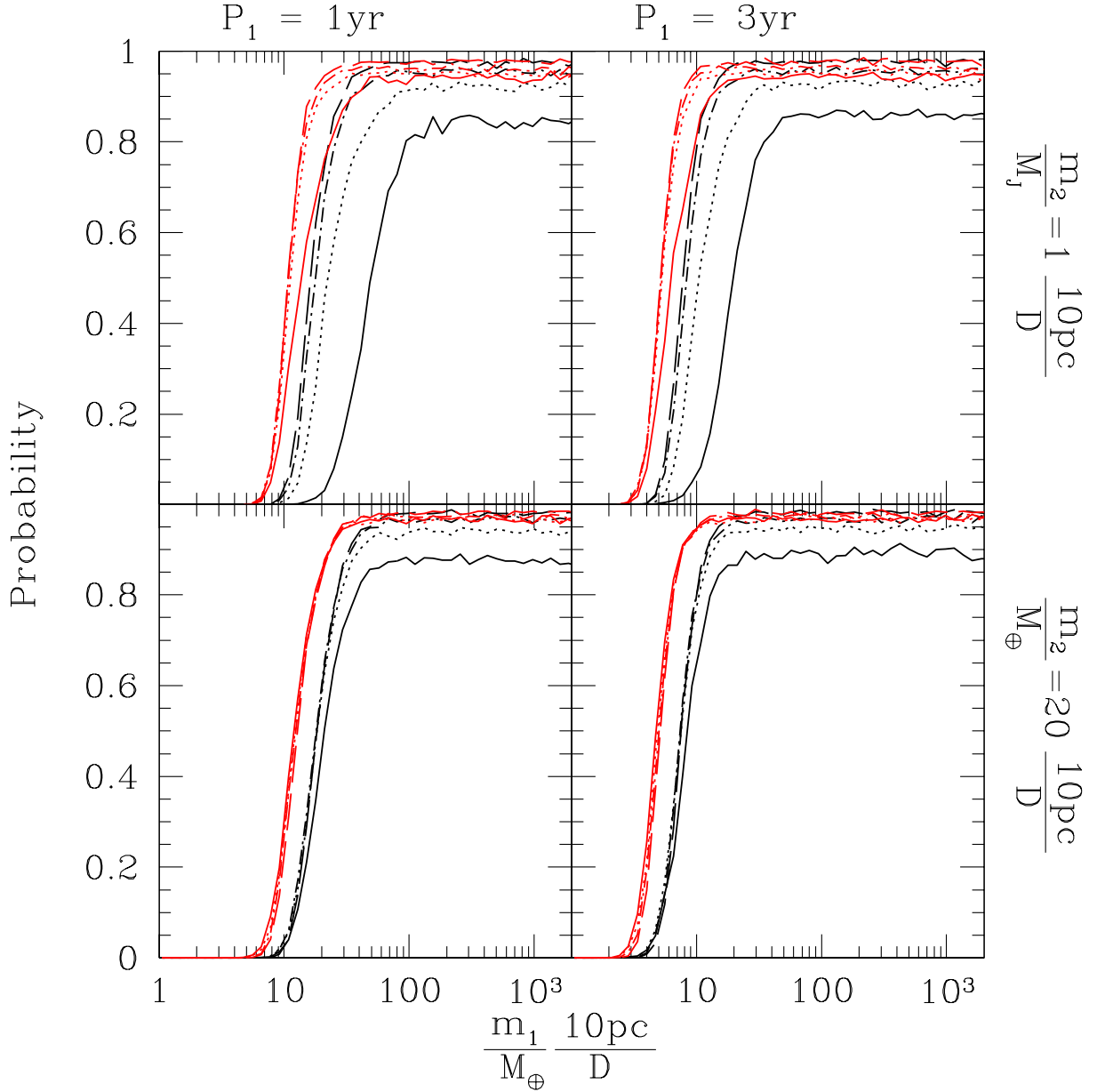}
\caption[f15.eps]{
Here we show how the probability of detecting a planet in the presence
of a second planet in a short-period orbit is improved by adding
radial velocity observations.  We plot the probability of detecting a
planet (1) in a 1yr (left panels) or 3yr (right panels) period orbit
as a function of its mass ($m_1$).  The second planet is in a 3d orbit
has a mass of $1M_J$ (top panels) or $20M_\oplus$ (bottom panels).
These results can be scaled to different planet masses and stellar
distances, as indicated by the x-axis label and the panel
labels on the right.  The solid lines a for a survey with only 24 2-d astrometric
observations.  The remaining lines are for a survey including 24 2-d
astrometric observations and 12 (dotted), 24 (dotted-dashed), or 48
(long dashed) radial velocity observations with 3m/s precision.  The
electronic version includes red curves that assume 48 2-d astrometric
observations over a 10 year mission.
\label{Fig15}}
\end{figure}

\begin{figure}[ht]
\plotone{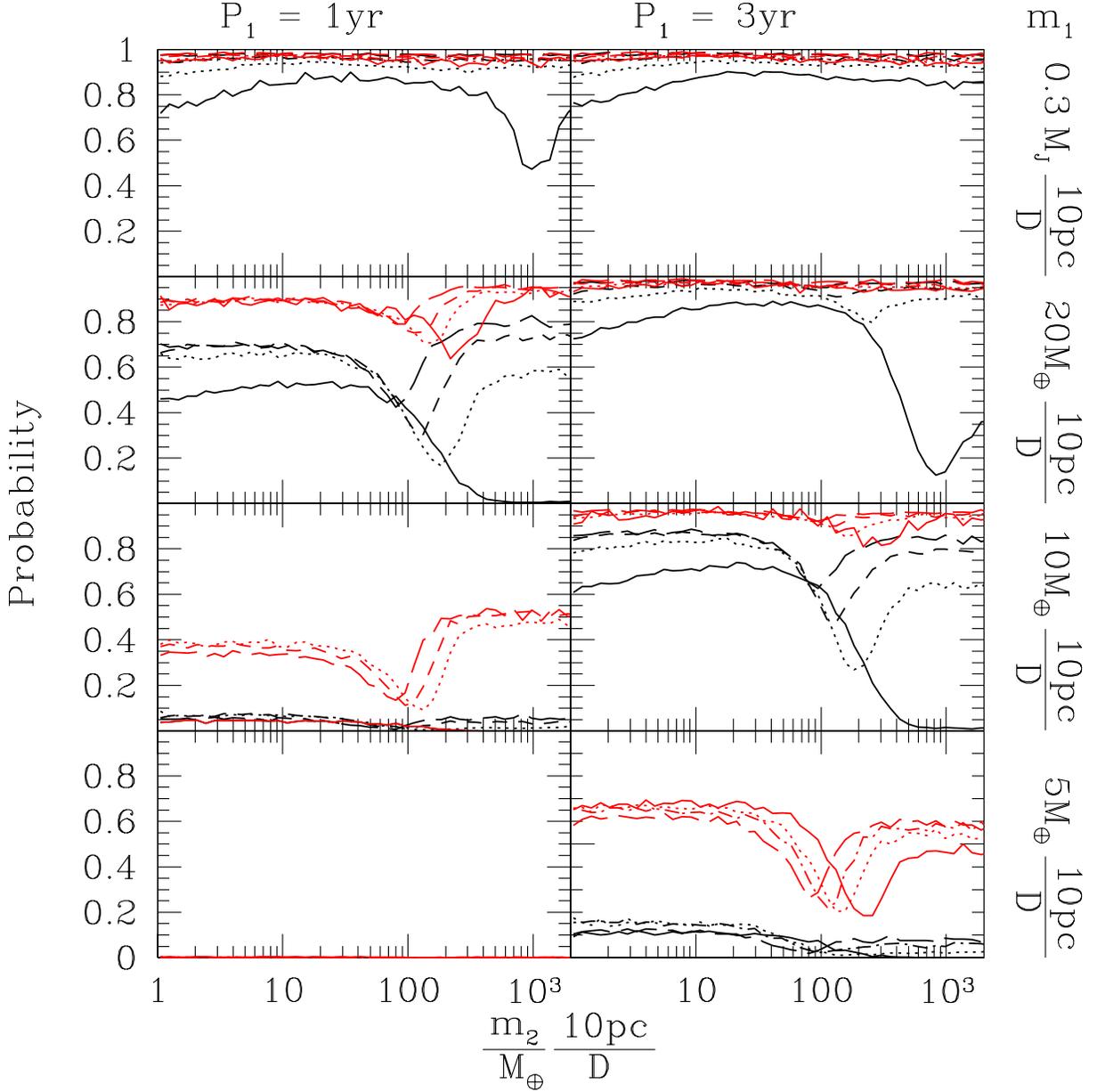}
\caption[f16.eps]{
Here we show how the probability of detecting a planet in the presence
of a second planet in a short-period orbit is improved by adding
radial velocity observations.  We plot the probability of detecting a
planet of mass $0.3M_J$ (top panels), $20M_\oplus$ (upper middle
panels), $10M_\oplus$ (lower middle panels), or $5M_\oplus$ (bottom
panels) in a 1yr (left panels) or 3yr (right panels) period orbit as a
function of the mass of the short-period planet ($m_2$) in a 3d orbit.
These results can be scaled to different planet masses and stellar
distances, as indicated by the x-axis label and the panel
labels on the right.  The solid lines a for a survey with only 24 2-d astrometric
observations.  The remaining lines are for a survey including 24 2-d
astrometric observations and 12 (dotted), 24 (dotted-dashed), or 48
(long dashed) radial velocity observations with 3m/s precision.  The
electronic version includes red curves that assume 48 2-d astrometric
observations over a 10 year mission.
\label{Fig16}}
\end{figure}


\clearpage

\begin{deluxetable}{cccc@{\qquad}cc@{\qquad}ccc@{\qquad}cc@{\qquad}ccc}
\tablecaption{} 
\tablehead{ \multicolumn{2}{c}{Mass} & \multicolumn{2}{c}{Period} & \multicolumn{2}{l}{One Planet} & \multicolumn{3}{l}{Two Planets} \\ A & B & A & B & A & B & A & B & A\&B } 
\startdata
5 $M_\oplus$ & 5 $M_\oplus$ & 3 d & 60 d &  0.00 & 0.00 & - & - & 0.00   \\
5 $M_\oplus$ & 10 $M_\oplus$ & 3 d & 60 d &  0.00 & 0.00 & - & - & 0.00   \\
5 $M_\oplus$ & 20 $M_\oplus$ & 3 d & 60 d &  0.00 & 0.00 & - & - & 0.00   \\
5 $M_\oplus$ & 0.3 $M_J$ & 3 d & 60 d &  0.00 & 0.91 & - & 0.82 & 0.00   \\
5 $M_\oplus$ & 1 $M_J$ & 3 d & 60 d &  0.00 & 1.00 & - & 0.81 & 0.00   \\
10 $M_\oplus$ & 5 $M_\oplus$ & 3 d & 60 d &  0.00 & 0.00 & - & - & 0.00   \\
10 $M_\oplus$ & 10 $M_\oplus$ & 3 d & 60 d &  0.00 & 0.00 & - & - & 0.00   \\
10 $M_\oplus$ & 20 $M_\oplus$ & 3 d & 60 d &  0.00 & 0.00 & - & - & 0.00   \\
10 $M_\oplus$ & 0.3 $M_J$ & 3 d & 60 d &  0.00 & 0.91 & - & 0.86 & 0.00   \\
10 $M_\oplus$ & 1 $M_J$ & 3 d & 60 d &  0.00 & 1.00 & - & 0.84 & 0.00   \\
20 $M_\oplus$ & 5 $M_\oplus$ & 3 d & 60 d &  0.00 & 0.00 & - & - & 0.00   \\
20 $M_\oplus$ & 10 $M_\oplus$ & 3 d & 60 d &  0.00 & 0.00 & - & - & 0.00   \\
20 $M_\oplus$ & 20 $M_\oplus$ & 3 d & 60 d &  0.00 & 0.00 & - & - & 0.00   \\
20 $M_\oplus$ & 0.3 $M_J$ & 3 d & 60 d &  0.00 & 0.91 & - & 0.87 & 0.00   \\
20 $M_\oplus$ & 1 $M_J$ & 3 d & 60 d &  0.00 & 1.00 & - & 0.86 & 0.00   \\
0.3 $M_J$ & 5 $M_\oplus$ & 3 d & 60 d &  0.00 & 0.00 & - & - & 0.00   \\
0.3 $M_J$ & 10 $M_\oplus$ & 3 d & 60 d &  0.00 & 0.00 & - & - & 0.00   \\
0.3 $M_J$ & 20 $M_\oplus$ & 3 d & 60 d &  0.00 & 0.00 & - & - & 0.00   \\
0.3 $M_J$ & 0.3 $M_J$ & 3 d & 60 d &  0.00 & 0.91 & - & 0.80 & 0.00   \\
0.3 $M_J$ & 1 $M_J$ & 3 d & 60 d &  0.00 & 1.00 & - & 0.84 & 0.00   \\
1 $M_J$ & 5 $M_\oplus$ & 3 d & 60 d &  0.11 & 0.00 & 0.87 & - & 0.00   \\
1 $M_J$ & 10 $M_\oplus$ & 3 d & 60 d &  0.11 & 0.00 & 0.62 & - & 0.00   \\
1 $M_J$ & 20 $M_\oplus$ & 3 d & 60 d &  0.11 & 0.00 & 0.16 & - & 0.00   \\
1 $M_J$ & 0.3 $M_J$ & 3 d & 60 d &  0.11 & 0.91 & 0.00 & 0.21 & 0.00   \\
1 $M_J$ & 1 $M_J$ & 3 d & 60 d &  0.11 & 1.00 & 0.00 & 0.82 & 0.00   \\
5 $M_\oplus$ & 5 $M_\oplus$ & 3 d & 1 yr &  0.00 & 0.00 & - & - & 0.00   \\
5 $M_\oplus$ & 10 $M_\oplus$ & 3 d & 1 yr &  0.00 & 0.04 & - & 0.98 & 0.00   \\
5 $M_\oplus$ & 20 $M_\oplus$ & 3 d & 1 yr &  0.00 & 0.58 & - & 0.90 & 0.00   \\
5 $M_\oplus$ & 0.3 $M_J$ & 3 d & 1 yr &  0.00 & 1.00 & - & 0.82 & 0.00   \\
5 $M_\oplus$ & 1 $M_J$ & 3 d & 1 yr &  0.00 & 1.00 & - & 0.83 & 0.00   \\
10 $M_\oplus$ & 5 $M_\oplus$ & 3 d & 1 yr &  0.00 & 0.00 & - & - & 0.00   \\
10 $M_\oplus$ & 10 $M_\oplus$ & 3 d & 1 yr &  0.00 & 0.04 & - & 0.97 & 0.00   \\
10 $M_\oplus$ & 20 $M_\oplus$ & 3 d & 1 yr &  0.00 & 0.58 & - & 0.91 & 0.00   \\
10 $M_\oplus$ & 0.3 $M_J$ & 3 d & 1 yr &  0.00 & 1.00 & - & 0.86 & 0.00   \\
10 $M_\oplus$ & 1 $M_J$ & 3 d & 1 yr &  0.00 & 1.00 & - & 0.86 & 0.00   \\
20 $M_\oplus$ & 5 $M_\oplus$ & 3 d & 1 yr &  0.00 & 0.00 & - & - & 0.00   \\
20 $M_\oplus$ & 10 $M_\oplus$ & 3 d & 1 yr &  0.00 & 0.04 & - & 0.97 & 0.00   \\
20 $M_\oplus$ & 20 $M_\oplus$ & 3 d & 1 yr &  0.00 & 0.58 & - & 0.90 & 0.00   \\
20 $M_\oplus$ & 0.3 $M_J$ & 3 d & 1 yr &  0.00 & 1.00 & - & 0.88 & 0.00   \\
20 $M_\oplus$ & 1 $M_J$ & 3 d & 1 yr &  0.00 & 1.00 & - & 0.88 & 0.00   \\
0.3 $M_J$ & 5 $M_\oplus$ & 3 d & 1 yr &  0.00 & 0.00 & - & - & 0.00   \\
0.3 $M_J$ & 10 $M_\oplus$ & 3 d & 1 yr &  0.00 & 0.04 & - & 0.51 & 0.00   \\
0.3 $M_J$ & 20 $M_\oplus$ & 3 d & 1 yr &  0.00 & 0.58 & - & 0.66 & 0.00   \\
0.3 $M_J$ & 0.3 $M_J$ & 3 d & 1 yr &  0.00 & 1.00 & - & 0.86 & 0.00   \\
0.3 $M_J$ & 1 $M_J$ & 3 d & 1 yr &  0.00 & 1.00 & - & 0.86 & 0.00   \\
1 $M_J$ & 5 $M_\oplus$ & 3 d & 1 yr &  0.11 & 0.00 & 0.22 & - & 0.00   \\
1 $M_J$ & 10 $M_\oplus$ & 3 d & 1 yr &  0.11 & 0.04 & 0.03 & 0.03 & 0.00   \\
1 $M_J$ & 20 $M_\oplus$ & 3 d & 1 yr &  0.11 & 0.58 & 0.00 & 0.07 & 0.00   \\
1 $M_J$ & 0.3 $M_J$ & 3 d & 1 yr &  0.11 & 1.00 & 0.00 & 0.81 & 0.00   \\
1 $M_J$ & 1 $M_J$ & 3 d & 1 yr &  0.11 & 1.00 & 0.00 & 0.85 & 0.00   \\
5 $M_\oplus$ & 5 $M_\oplus$ & 3 d & 3 yr &  0.00 & 0.12 & - & 0.93 & 0.00   \\
5 $M_\oplus$ & 10 $M_\oplus$ & 3 d & 3 yr &  0.00 & 0.79 & - & 0.87 & 0.00   \\
5 $M_\oplus$ & 20 $M_\oplus$ & 3 d & 3 yr &  0.00 & 0.99 & - & 0.85 & 0.00   \\
5 $M_\oplus$ & 0.3 $M_J$ & 3 d & 3 yr &  0.00 & 1.00 & - & 0.84 & 0.00   \\
5 $M_\oplus$ & 1 $M_J$ & 3 d & 3 yr &  0.00 & 1.00 & - & 0.85 & 0.00   \\
10 $M_\oplus$ & 5 $M_\oplus$ & 3 d & 3 yr &  0.00 & 0.12 & - & 0.97 & 0.00   \\
10 $M_\oplus$ & 10 $M_\oplus$ & 3 d & 3 yr &  0.00 & 0.79 & - & 0.91 & 0.00   \\
10 $M_\oplus$ & 20 $M_\oplus$ & 3 d & 3 yr &  0.00 & 0.99 & - & 0.88 & 0.00   \\
10 $M_\oplus$ & 0.3 $M_J$ & 3 d & 3 yr &  0.00 & 1.00 & - & 0.88 & 0.00   \\
10 $M_\oplus$ & 1 $M_J$ & 3 d & 3 yr &  0.00 & 1.00 & - & 0.87 & 0.00   \\
20 $M_\oplus$ & 5 $M_\oplus$ & 3 d & 3 yr &  0.00 & 0.12 & - & 0.96 & 0.00   \\
20 $M_\oplus$ & 10 $M_\oplus$ & 3 d & 3 yr &  0.00 & 0.79 & - & 0.92 & 0.00   \\
20 $M_\oplus$ & 20 $M_\oplus$ & 3 d & 3 yr &  0.00 & 0.99 & - & 0.89 & 0.00   \\
20 $M_\oplus$ & 0.3 $M_J$ & 3 d & 3 yr &  0.00 & 1.00 & - & 0.89 & 0.00   \\
20 $M_\oplus$ & 1 $M_J$ & 3 d & 3 yr &  0.00 & 1.00 & - & 0.89 & 0.00   \\
0.3 $M_J$ & 5 $M_\oplus$ & 3 d & 3 yr &  0.00 & 0.12 & - & 0.47 & 0.00   \\
0.3 $M_J$ & 10 $M_\oplus$ & 3 d & 3 yr &  0.00 & 0.79 & - & 0.75 & 0.00   \\
0.3 $M_J$ & 20 $M_\oplus$ & 3 d & 3 yr &  0.00 & 0.99 & - & 0.87 & 0.00   \\
0.3 $M_J$ & 0.3 $M_J$ & 3 d & 3 yr &  0.00 & 1.00 & - & 0.88 & 0.00   \\
0.3 $M_J$ & 1 $M_J$ & 3 d & 3 yr &  0.00 & 1.00 & - & 0.88 & 0.00   \\
1 $M_J$ & 5 $M_\oplus$ & 3 d & 3 yr &  0.11 & 0.12 & 0.02 & 0.02 & 0.00   \\
1 $M_J$ & 10 $M_\oplus$ & 3 d & 3 yr &  0.11 & 0.79 & 0.00 & 0.11 & 0.00   \\
1 $M_J$ & 20 $M_\oplus$ & 3 d & 3 yr &  0.11 & 0.99 & 0.00 & 0.59 & 0.00   \\
1 $M_J$ & 0.3 $M_J$ & 3 d & 3 yr &  0.11 & 1.00 & 0.00 & 0.85 & 0.00   \\
1 $M_J$ & 1 $M_J$ & 3 d & 3 yr &  0.11 & 1.00 & 0.00 & 0.87 & 0.00   \\
5 $M_\oplus$ & 5 $M_\oplus$ & 3 d & 12 yr &  0.00 & 0.04 & - & 0.91 & 0.00   \\
5 $M_\oplus$ & 10 $M_\oplus$ & 3 d & 12 yr &  0.00 & 0.31 & - & 0.87 & 0.00   \\
5 $M_\oplus$ & 20 $M_\oplus$ & 3 d & 12 yr &  0.00 & 0.68 & - & 0.79 & 0.00   \\
5 $M_\oplus$ & 0.3 $M_J$ & 3 d & 12 yr &  0.00 & 0.99 & - & 0.75 & 0.00   \\
5 $M_\oplus$ & 1 $M_J$ & 3 d & 12 yr &  0.00 & 1.00 & - & 0.74 & 0.00   \\
10 $M_\oplus$ & 5 $M_\oplus$ & 3 d & 12 yr &  0.00 & 0.04 & - & 0.99 & 0.00   \\
10 $M_\oplus$ & 10 $M_\oplus$ & 3 d & 12 yr &  0.00 & 0.31 & - & 0.88 & 0.00   \\
10 $M_\oplus$ & 20 $M_\oplus$ & 3 d & 12 yr &  0.00 & 0.68 & - & 0.82 & 0.00   \\
10 $M_\oplus$ & 0.3 $M_J$ & 3 d & 12 yr &  0.00 & 0.99 & - & 0.77 & 0.00   \\
10 $M_\oplus$ & 1 $M_J$ & 3 d & 12 yr &  0.00 & 1.00 & - & 0.77 & 0.00   \\
20 $M_\oplus$ & 5 $M_\oplus$ & 3 d & 12 yr &  0.00 & 0.04 & - & 0.90 & 0.00   \\
20 $M_\oplus$ & 10 $M_\oplus$ & 3 d & 12 yr &  0.00 & 0.31 & - & 0.88 & 0.00   \\
20 $M_\oplus$ & 20 $M_\oplus$ & 3 d & 12 yr &  0.00 & 0.68 & - & 0.83 & 0.00   \\
20 $M_\oplus$ & 0.3 $M_J$ & 3 d & 12 yr &  0.00 & 0.99 & - & 0.78 & 0.00   \\
20 $M_\oplus$ & 1 $M_J$ & 3 d & 12 yr &  0.00 & 1.00 & - & 0.78 & 0.00   \\
0.3 $M_J$ & 5 $M_\oplus$ & 3 d & 12 yr &  0.00 & 0.04 & - & 0.56 & 0.00   \\
0.3 $M_J$ & 10 $M_\oplus$ & 3 d & 12 yr &  0.00 & 0.31 & - & 0.68 & 0.00   \\
0.3 $M_J$ & 20 $M_\oplus$ & 3 d & 12 yr &  0.00 & 0.68 & - & 0.74 & 0.00   \\
0.3 $M_J$ & 0.3 $M_J$ & 3 d & 12 yr &  0.00 & 0.99 & - & 0.77 & 0.00   \\
0.3 $M_J$ & 1 $M_J$ & 3 d & 12 yr &  0.00 & 1.00 & - & 0.77 & 0.00   \\
1 $M_J$ & 5 $M_\oplus$ & 3 d & 12 yr &  0.11 & 0.04 & 0.19 & 0.03 & 0.00   \\
1 $M_J$ & 10 $M_\oplus$ & 3 d & 12 yr &  0.11 & 0.31 & 0.05 & 0.11 & 0.00   \\
1 $M_J$ & 20 $M_\oplus$ & 3 d & 12 yr &  0.11 & 0.68 & 0.00 & 0.32 & 0.00   \\
1 $M_J$ & 0.3 $M_J$ & 3 d & 12 yr &  0.11 & 0.99 & 0.00 & 0.71 & 0.00   \\
1 $M_J$ & 1 $M_J$ & 3 d & 12 yr &  0.11 & 1.00 & 0.00 & 0.76 & 0.00   \\
5 $M_\oplus$ & 5 $M_\oplus$ & 60 d & 1 yr &  0.00 & 0.00 & - & - & 0.00   \\
5 $M_\oplus$ & 10 $M_\oplus$ & 60 d & 1 yr &  0.00 & 0.04 & - & 0.86 & 0.00   \\
5 $M_\oplus$ & 20 $M_\oplus$ & 60 d & 1 yr &  0.00 & 0.58 & - & 0.92 & 0.00   \\
5 $M_\oplus$ & 0.3 $M_J$ & 60 d & 1 yr &  0.00 & 1.00 & - & 0.94 & 0.00   \\
5 $M_\oplus$ & 1 $M_J$ & 60 d & 1 yr &  0.00 & 1.00 & - & 0.95 & 0.00   \\
10 $M_\oplus$ & 5 $M_\oplus$ & 60 d & 1 yr &  0.00 & 0.00 & - & - & 0.00   \\
10 $M_\oplus$ & 10 $M_\oplus$ & 60 d & 1 yr &  0.00 & 0.04 & - & 0.63 & 0.00   \\
10 $M_\oplus$ & 20 $M_\oplus$ & 60 d & 1 yr &  0.00 & 0.58 & - & 0.82 & 0.00   \\
10 $M_\oplus$ & 0.3 $M_J$ & 60 d & 1 yr &  0.00 & 1.00 & - & 0.96 & 0.00   \\
10 $M_\oplus$ & 1 $M_J$ & 60 d & 1 yr &  0.00 & 1.00 & - & 0.96 & 0.00   \\
20 $M_\oplus$ & 5 $M_\oplus$ & 60 d & 1 yr &  0.00 & 0.00 & - & - & 0.00   \\
20 $M_\oplus$ & 10 $M_\oplus$ & 60 d & 1 yr &  0.00 & 0.04 & - & 0.22 & 0.00   \\
20 $M_\oplus$ & 20 $M_\oplus$ & 60 d & 1 yr &  0.00 & 0.58 & - & 0.50 & 0.00   \\
20 $M_\oplus$ & 0.3 $M_J$ & 60 d & 1 yr &  0.00 & 1.00 & - & 0.96 & 0.00   \\
20 $M_\oplus$ & 1 $M_J$ & 60 d & 1 yr &  0.00 & 1.00 & - & 0.96 & 0.00   \\
0.3 $M_J$ & 5 $M_\oplus$ & 60 d & 1 yr &  0.91 & 0.00 & 0.85 & - & 0.00   \\
0.3 $M_J$ & 10 $M_\oplus$ & 60 d & 1 yr &  0.91 & 0.04 & 0.52 & 0.01 & 0.00   \\
0.3 $M_J$ & 20 $M_\oplus$ & 60 d & 1 yr &  0.91 & 0.58 & 0.12 & 0.04 & 0.02   \\
0.3 $M_J$ & 0.3 $M_J$ & 60 d & 1 yr &  0.91 & 1.00 & 0.16 & 0.72 & 0.15   \\
0.3 $M_J$ & 1 $M_J$ & 60 d & 1 yr &  0.91 & 1.00 & 0.17 & 0.98 & 0.15   \\
1 $M_J$ & 5 $M_\oplus$ & 60 d & 1 yr &  1.00 & 0.00 & 0.96 & - & 0.00   \\
1 $M_J$ & 10 $M_\oplus$ & 60 d & 1 yr &  1.00 & 0.04 & 0.96 & 0.01 & 0.00   \\
1 $M_J$ & 20 $M_\oplus$ & 60 d & 1 yr &  1.00 & 0.58 & 0.92 & 0.03 & 0.02   \\
1 $M_J$ & 0.3 $M_J$ & 60 d & 1 yr &  1.00 & 1.00 & 0.94 & 0.92 & 0.91   \\
1 $M_J$ & 1 $M_J$ & 60 d & 1 yr &  1.00 & 1.00 & 0.96 & 0.98 & 0.95   \\
5 $M_\oplus$ & 5 $M_\oplus$ & 60 d & 3 yr &  0.00 & 0.12 & - & 0.88 & 0.00   \\
5 $M_\oplus$ & 10 $M_\oplus$ & 60 d & 3 yr &  0.00 & 0.79 & - & 0.95 & 0.00   \\
5 $M_\oplus$ & 20 $M_\oplus$ & 60 d & 3 yr &  0.00 & 0.99 & - & 0.95 & 0.00   \\
5 $M_\oplus$ & 0.3 $M_J$ & 60 d & 3 yr &  0.00 & 1.00 & - & 0.96 & 0.00   \\
5 $M_\oplus$ & 1 $M_J$ & 60 d & 3 yr &  0.00 & 1.00 & - & 0.95 & 0.00   \\
10 $M_\oplus$ & 5 $M_\oplus$ & 60 d & 3 yr &  0.00 & 0.12 & - & 0.63 & 0.00   \\
10 $M_\oplus$ & 10 $M_\oplus$ & 60 d & 3 yr &  0.00 & 0.79 & - & 0.88 & 0.00   \\
10 $M_\oplus$ & 20 $M_\oplus$ & 60 d & 3 yr &  0.00 & 0.99 & - & 0.95 & 0.00   \\
10 $M_\oplus$ & 0.3 $M_J$ & 60 d & 3 yr &  0.00 & 1.00 & - & 0.96 & 0.00   \\
10 $M_\oplus$ & 1 $M_J$ & 60 d & 3 yr &  0.00 & 1.00 & - & 0.97 & 0.00   \\
20 $M_\oplus$ & 5 $M_\oplus$ & 60 d & 3 yr &  0.00 & 0.12 & - & 0.21 & 0.00   \\
20 $M_\oplus$ & 10 $M_\oplus$ & 60 d & 3 yr &  0.00 & 0.79 & - & 0.62 & 0.00   \\
20 $M_\oplus$ & 20 $M_\oplus$ & 60 d & 3 yr &  0.00 & 0.99 & - & 0.94 & 0.00   \\
20 $M_\oplus$ & 0.3 $M_J$ & 60 d & 3 yr &  0.00 & 1.00 & - & 0.97 & 0.00   \\
20 $M_\oplus$ & 1 $M_J$ & 60 d & 3 yr &  0.00 & 1.00 & - & 0.98 & 0.00   \\
0.3 $M_J$ & 5 $M_\oplus$ & 60 d & 3 yr &  0.91 & 0.12 & 0.41 & 0.02 & 0.00   \\
0.3 $M_J$ & 10 $M_\oplus$ & 60 d & 3 yr &  0.91 & 0.79 & 0.10 & 0.07 & 0.05   \\
0.3 $M_J$ & 20 $M_\oplus$ & 60 d & 3 yr &  0.91 & 0.99 & 0.18 & 0.27 & 0.16   \\
0.3 $M_J$ & 0.3 $M_J$ & 60 d & 3 yr &  0.91 & 1.00 & 0.20 & 0.98 & 0.18   \\
0.3 $M_J$ & 1 $M_J$ & 60 d & 3 yr &  0.91 & 1.00 & 0.19 & 0.99 & 0.18   \\
1 $M_J$ & 5 $M_\oplus$ & 60 d & 3 yr &  1.00 & 0.12 & 0.97 & 0.02 & 0.00   \\
1 $M_J$ & 10 $M_\oplus$ & 60 d & 3 yr &  1.00 & 0.79 & 0.91 & 0.08 & 0.06   \\
1 $M_J$ & 20 $M_\oplus$ & 60 d & 3 yr &  1.00 & 0.99 & 0.82 & 0.59 & 0.59   \\
1 $M_J$ & 0.3 $M_J$ & 60 d & 3 yr &  1.00 & 1.00 & 0.96 & 0.98 & 0.96   \\
1 $M_J$ & 1 $M_J$ & 60 d & 3 yr &  1.00 & 1.00 & 0.96 & 0.99 & 0.96   \\
5 $M_\oplus$ & 5 $M_\oplus$ & 60 d & 12 yr &  0.00 & 0.04 & - & 0.89 & 0.00   \\
5 $M_\oplus$ & 10 $M_\oplus$ & 60 d & 12 yr &  0.00 & 0.31 & - & 0.91 & 0.00   \\
5 $M_\oplus$ & 20 $M_\oplus$ & 60 d & 12 yr &  0.00 & 0.68 & - & 0.89 & 0.00   \\
5 $M_\oplus$ & 0.3 $M_J$ & 60 d & 12 yr &  0.00 & 0.99 & - & 0.89 & 0.00   \\
5 $M_\oplus$ & 1 $M_J$ & 60 d & 12 yr &  0.00 & 1.00 & - & 0.89 & 0.00   \\
10 $M_\oplus$ & 5 $M_\oplus$ & 60 d & 12 yr &  0.00 & 0.04 & - & 0.70 & 0.00   \\
10 $M_\oplus$ & 10 $M_\oplus$ & 60 d & 12 yr &  0.00 & 0.31 & - & 0.81 & 0.00   \\
10 $M_\oplus$ & 20 $M_\oplus$ & 60 d & 12 yr &  0.00 & 0.68 & - & 0.88 & 0.00   \\
10 $M_\oplus$ & 0.3 $M_J$ & 60 d & 12 yr &  0.00 & 0.99 & - & 0.90 & 0.00   \\
10 $M_\oplus$ & 1 $M_J$ & 60 d & 12 yr &  0.00 & 1.00 & - & 0.91 & 0.00   \\
20 $M_\oplus$ & 5 $M_\oplus$ & 60 d & 12 yr &  0.00 & 0.04 & - & 0.31 & 0.00   \\
20 $M_\oplus$ & 10 $M_\oplus$ & 60 d & 12 yr &  0.00 & 0.31 & - & 0.54 & 0.00   \\
20 $M_\oplus$ & 20 $M_\oplus$ & 60 d & 12 yr &  0.00 & 0.68 & - & 0.75 & 0.00   \\
20 $M_\oplus$ & 0.3 $M_J$ & 60 d & 12 yr &  0.00 & 0.99 & - & 0.92 & 0.00   \\
20 $M_\oplus$ & 1 $M_J$ & 60 d & 12 yr &  0.00 & 1.00 & - & 0.93 & 0.00   \\
0.3 $M_J$ & 5 $M_\oplus$ & 60 d & 12 yr &  0.91 & 0.04 & 0.68 & 0.03 & 0.00   \\
0.3 $M_J$ & 10 $M_\oplus$ & 60 d & 12 yr &  0.91 & 0.31 & 0.37 & 0.05 & 0.01   \\
0.3 $M_J$ & 20 $M_\oplus$ & 60 d & 12 yr &  0.91 & 0.68 & 0.18 & 0.15 & 0.06   \\
0.3 $M_J$ & 0.3 $M_J$ & 60 d & 12 yr &  0.91 & 0.99 & 0.13 & 0.75 & 0.12   \\
0.3 $M_J$ & 1 $M_J$ & 60 d & 12 yr &  0.91 & 1.00 & 0.13 & 0.94 & 0.12   \\
1 $M_J$ & 5 $M_\oplus$ & 60 d & 12 yr &  1.00 & 0.04 & 0.94 & 0.03 & 0.00   \\
1 $M_J$ & 10 $M_\oplus$ & 60 d & 12 yr &  1.00 & 0.31 & 0.92 & 0.07 & 0.02   \\
1 $M_J$ & 20 $M_\oplus$ & 60 d & 12 yr &  1.00 & 0.68 & 0.85 & 0.28 & 0.19   \\
1 $M_J$ & 0.3 $M_J$ & 60 d & 12 yr &  1.00 & 0.99 & 0.90 & 0.88 & 0.85   \\
1 $M_J$ & 1 $M_J$ & 60 d & 12 yr &  1.00 & 1.00 & 0.94 & 0.96 & 0.93   \\
5 $M_\oplus$ & 5 $M_\oplus$ & 1 yr & 3 yr &  0.00 & 0.12 & - & 0.37 & 0.00   \\
5 $M_\oplus$ & 10 $M_\oplus$ & 1 yr & 3 yr &  0.00 & 0.79 & - & 0.73 & 0.00   \\
5 $M_\oplus$ & 20 $M_\oplus$ & 1 yr & 3 yr &  0.00 & 0.99 & - & 0.94 & 0.00   \\
5 $M_\oplus$ & 0.3 $M_J$ & 1 yr & 3 yr &  0.00 & 1.00 & - & 0.95 & 0.00   \\
5 $M_\oplus$ & 1 $M_J$ & 1 yr & 3 yr &  0.00 & 1.00 & - & 0.96 & 0.00   \\
10 $M_\oplus$ & 5 $M_\oplus$ & 1 yr & 3 yr &  0.04 & 0.12 & 0.04 & 0.08 & 0.00   \\
10 $M_\oplus$ & 10 $M_\oplus$ & 1 yr & 3 yr &  0.04 & 0.79 & 0.06 & 0.31 & 0.00   \\
10 $M_\oplus$ & 20 $M_\oplus$ & 1 yr & 3 yr &  0.04 & 0.99 & 0.12 & 0.85 & 0.00   \\
10 $M_\oplus$ & 0.3 $M_J$ & 1 yr & 3 yr &  0.04 & 1.00 & 0.49 & 0.96 & 0.00   \\
10 $M_\oplus$ & 1 $M_J$ & 1 yr & 3 yr &  0.04 & 1.00 & 0.50 & 0.96 & 0.00   \\
20 $M_\oplus$ & 5 $M_\oplus$ & 1 yr & 3 yr &  0.58 & 0.12 & 0.11 & 0.03 & 0.00   \\
20 $M_\oplus$ & 10 $M_\oplus$ & 1 yr & 3 yr &  0.58 & 0.79 & 0.07 & 0.09 & 0.04   \\
20 $M_\oplus$ & 20 $M_\oplus$ & 1 yr & 3 yr &  0.58 & 0.99 & 0.10 & 0.43 & 0.06   \\
20 $M_\oplus$ & 0.3 $M_J$ & 1 yr & 3 yr &  0.58 & 1.00 & 0.11 & 0.98 & 0.05   \\
20 $M_\oplus$ & 1 $M_J$ & 1 yr & 3 yr &  0.58 & 1.00 & 0.12 & 0.97 & 0.05   \\
0.3 $M_J$ & 5 $M_\oplus$ & 1 yr & 3 yr &  1.00 & 0.12 & 0.96 & 0.04 & 0.00   \\
0.3 $M_J$ & 10 $M_\oplus$ & 1 yr & 3 yr &  1.00 & 0.79 & 0.84 & 0.14 & 0.11   \\
0.3 $M_J$ & 20 $M_\oplus$ & 1 yr & 3 yr &  1.00 & 0.99 & 0.83 & 0.68 & 0.67   \\
0.3 $M_J$ & 0.3 $M_J$ & 1 yr & 3 yr &  1.00 & 1.00 & 0.95 & 0.99 & 0.95   \\
0.3 $M_J$ & 1 $M_J$ & 1 yr & 3 yr &  1.00 & 1.00 & 0.95 & 0.99 & 0.94   \\
1 $M_J$ & 5 $M_\oplus$ & 1 yr & 3 yr &  1.00 & 0.12 & 0.98 & 0.04 & 0.00   \\
1 $M_J$ & 10 $M_\oplus$ & 1 yr & 3 yr &  1.00 & 0.79 & 0.98 & 0.15 & 0.11   \\
1 $M_J$ & 20 $M_\oplus$ & 1 yr & 3 yr &  1.00 & 0.99 & 0.99 & 0.69 & 0.69   \\
1 $M_J$ & 0.3 $M_J$ & 1 yr & 3 yr &  1.00 & 1.00 & 1.00 & 0.99 & 0.99   \\
1 $M_J$ & 1 $M_J$ & 1 yr & 3 yr &  1.00 & 1.00 & 1.00 & 1.00 & 1.00   \\
5 $M_\oplus$ & 5 $M_\oplus$ & 1 yr & 12 yr &  0.00 & 0.04 & - & 0.35 & 0.00   \\
5 $M_\oplus$ & 10 $M_\oplus$ & 1 yr & 12 yr &  0.00 & 0.31 & - & 0.64 & 0.00   \\
5 $M_\oplus$ & 0.3 $M_J$ & 1 yr & 12 yr &  0.00 & 0.99 & - & 0.91 & 0.00   \\
10 $M_\oplus$ & 5 $M_\oplus$ & 1 yr & 12 yr &  0.04 & 0.04 & 0.27 & 0.09 & 0.00   \\
10 $M_\oplus$ & 10 $M_\oplus$ & 1 yr & 12 yr &  0.04 & 0.31 & 0.12 & 0.26 & 0.00   \\
10 $M_\oplus$ & 0.3 $M_J$ & 1 yr & 12 yr &  0.04 & 0.99 & 0.52 & 0.91 & 0.00   \\
20 $M_\oplus$ & 5 $M_\oplus$ & 1 yr & 12 yr &  0.58 & 0.04 & 0.43 & 0.01 & 0.00   \\
20 $M_\oplus$ & 10 $M_\oplus$ & 1 yr & 12 yr &  0.58 & 0.31 & 0.17 & 0.07 & 0.01   \\
20 $M_\oplus$ & 20 $M_\oplus$ & 1 yr & 12 yr &  0.58 & 0.68 & 0.08 & 0.23 & 0.02   \\
20 $M_\oplus$ & 0.3 $M_J$ & 1 yr & 12 yr &  0.58 & 0.99 & 0.09 & 0.84 & 0.03   \\
20 $M_\oplus$ & 1 $M_J$ & 1 yr & 12 yr &  0.58 & 1.00 & 0.07 & 0.96 & 0.03   \\
0.3 $M_J$ & 5 $M_\oplus$ & 1 yr & 12 yr &  1.00 & 0.04 & 0.95 & 0.05 & 0.00   \\
0.3 $M_J$ & 10 $M_\oplus$ & 1 yr & 12 yr &  1.00 & 0.31 & 0.90 & 0.14 & 0.04   \\
0.3 $M_J$ & 0.3 $M_J$ & 1 yr & 12 yr &  1.00 & 0.99 & 0.92 & 0.91 & 0.88   \\
1 $M_J$ & 5 $M_\oplus$ & 1 yr & 12 yr &  1.00 & 0.04 & 0.96 & 0.06 & 0.00   \\
1 $M_J$ & 10 $M_\oplus$ & 1 yr & 12 yr &  1.00 & 0.31 & 0.97 & 0.13 & 0.04   \\
1 $M_J$ & 20 $M_\oplus$ & 1 yr & 12 yr &  1.00 & 0.68 & 0.98 & 0.37 & 0.25   \\
1 $M_J$ & 0.3 $M_J$ & 1 yr & 12 yr &  1.00 & 0.99 & 0.99 & 0.92 & 0.91   \\
1 $M_J$ & 1 $M_J$ & 1 yr & 12 yr &  1.00 & 1.00 & 0.99 & 0.98 & 0.98   \\
5 $M_\oplus$ & 5 $M_\oplus$ & 3 yr & 12 yr &  0.12 & 0.04 & 0.72 & 0.81 & 0.00   \\
5 $M_\oplus$ & 10 $M_\oplus$ & 3 yr & 12 yr &  0.12 & 0.31 & 0.99 & 0.47 & 0.00   \\
5 $M_\oplus$ & 20 $M_\oplus$ & 3 yr & 12 yr &  0.12 & 0.68 & 1.42 & 0.50 & 0.00   \\
5 $M_\oplus$ & 0.3 $M_J$ & 3 yr & 12 yr &  0.12 & 0.99 & 1.12 & 0.81 & 0.00   \\
5 $M_\oplus$ & 1 $M_J$ & 3 yr & 12 yr &  0.12 & 1.00 & 0.41 & 0.91 & 0.00   \\
10 $M_\oplus$ & 5 $M_\oplus$ & 3 yr & 12 yr &  0.79 & 0.04 & 0.66 & 1.37 & 0.00   \\
10 $M_\oplus$ & 10 $M_\oplus$ & 3 yr & 12 yr &  0.79 & 0.31 & 0.44 & 0.30 & 0.00   \\
10 $M_\oplus$ & 20 $M_\oplus$ & 3 yr & 12 yr &  0.79 & 0.68 & 0.32 & 0.32 & 0.01   \\
10 $M_\oplus$ & 0.3 $M_J$ & 3 yr & 12 yr &  0.79 & 0.99 & 0.24 & 0.71 & 0.01   \\
10 $M_\oplus$ & 1 $M_J$ & 3 yr & 12 yr &  0.79 & 1.00 & 0.11 & 0.89 & 0.01   \\
20 $M_\oplus$ & 5 $M_\oplus$ & 3 yr & 12 yr &  0.99 & 0.04 & 0.85 & 1.30 & 0.00   \\
20 $M_\oplus$ & 10 $M_\oplus$ & 3 yr & 12 yr &  0.99 & 0.31 & 0.74 & 0.25 & 0.01   \\
20 $M_\oplus$ & 20 $M_\oplus$ & 3 yr & 12 yr &  0.99 & 0.68 & 0.54 & 0.23 & 0.05   \\
20 $M_\oplus$ & 0.3 $M_J$ & 3 yr & 12 yr &  0.99 & 0.99 & 0.41 & 0.63 & 0.20   \\
20 $M_\oplus$ & 1 $M_J$ & 3 yr & 12 yr &  0.99 & 1.00 & 0.30 & 0.84 & 0.18   \\
0.3 $M_J$ & 5 $M_\oplus$ & 3 yr & 12 yr &  1.00 & 0.04 & 0.93 & 0.87 & 0.00   \\
0.3 $M_J$ & 10 $M_\oplus$ & 3 yr & 12 yr &  1.00 & 0.31 & 0.93 & 0.19 & 0.02   \\
0.3 $M_J$ & 20 $M_\oplus$ & 3 yr & 12 yr &  1.00 & 0.68 & 0.93 & 0.25 & 0.13   \\
0.3 $M_J$ & 0.3 $M_J$ & 3 yr & 12 yr &  1.00 & 0.99 & 0.97 & 0.80 & 0.77   \\
0.3 $M_J$ & 1 $M_J$ & 3 yr & 12 yr &  1.00 & 1.00 & 0.98 & 0.96 & 0.95   \\
1 $M_J$ & 5 $M_\oplus$ & 3 yr & 12 yr &  1.00 & 0.04 & 0.96 & 0.16 & 0.00   \\
1 $M_J$ & 10 $M_\oplus$ & 3 yr & 12 yr &  1.00 & 0.31 & 0.96 & 0.10 & 0.01   \\
1 $M_J$ & 20 $M_\oplus$ & 3 yr & 12 yr &  1.00 & 0.68 & 0.96 & 0.21 & 0.12   \\
1 $M_J$ & 0.3 $M_J$ & 3 yr & 12 yr &  1.00 & 0.99 & 0.98 & 0.81 & 0.80   \\
1 $M_J$ & 1 $M_J$ & 3 yr & 12 yr &  1.00 & 1.00 & 0.99 & 0.97 & 0.97   \\
\enddata
\tablecomments{}
\end{deluxetable}

\end{document}